\documentclass[a4paper,fleqn]{cas-sc}

\usepackage[numbers,sort&compress]{natbib}
\usepackage{grffile}

\def\tsc#1{\csdef{#1}{\textsc{\lowercase{#1}}\xspace}}
\tsc{WGM}
\tsc{QE}
\tsc{EP}
\tsc{PMS}
\tsc{BEC}
\tsc{DE}

\begin{document}

\let\WriteBookmarks\relax
\def\floatpagepagefraction{1}
\def\textpagefraction{.001}
\shorttitle{A numerical method for dynamic wetting}
\shortauthors{X.H. Wu et~al.}

\title [mode = title]{Macroscopic simulations of thin film wetting/dewetting using a precursor film model \\ }

\author[1]{Xi-Hu Wu}[style=chinese,orcid=0000-0002-8019-9384]
\affiliation[1]{
organization=
  {State Key Laboratory of High Temperature Gas Dynamics, School of Engineering Science, University of Science and Technology of China, Hefei, Anhui 230026, PR China}
}

\author[1]{Zhuo Long}[style=chinese]
\author[1]{Xue-Li Wang}[style=chinese]

\author[1]{Peng Gao}[style=chinese]
\cormark[1]
\ead{gaopeng@ustc.edu.cn}

\cortext[cor1]{Corresponding author}

\begin{abstract}
Numerical simulation of dynamic wetting processes remains challenging due to the multiscale feature of the moving contact line. This paper presents a finite element method for two- or three-dimensional dynamic wetting problems within the framework of the lubrication equation and the precursor film model. By reconstructing the disjoining pressure formulation, a mesoscopic precursor film model is implemented to enable a relatively thick precursor film to reproduce the intermediate-region interfacial behavior of a physically thin precursor film. The proposed model reduces the requirement for excessive spatial resolution near the contact line, thereby significantly reducing the computational cost while preserving the macroscopic flow dynamics. This method does not incorporate any moving boundary and easily handles topological changes. Its capability and accuracy are validated through various simulations, including the spreading, retraction, sliding, and coalescence of drops, as well as the breakup of liquid ridges. Numerical results (with the mesoscopic precursor film model adopted) show good agreement with the available exact solutions and asymptotic theories.
\end{abstract}




\maketitle

\section{Introduction}

The evolution of free surfaces and contact lines in wetting processes is crucial to various industrial applications~\cite{Kumar2015,Lohse2022,He2020}. In moving contact line problems, a purely hydrodynamic description with the conventional no-slip boundary condition induces a singularity in both viscous stress and pressure at the contact line, posing challenges for both numerical simulations and theoretical analyses~\cite{huh1971,snoeijer2013}. Researchers have proposed various microscopic contact line models to regularize this singularity, such as the slip model and the precursor film model. The widely adopted Navier slip model assumes a slip velocity at the solid wall that is proportional to the local shear rate via a nanoscale length, termed the slip length~\cite{huh1971}; the precursor film model assumes the existence of an ultra-thin liquid film (characterized by a nanoscale equilibrium film thickness) surrounding the macroscopic bulk liquid~\cite{de1985}. For numerical simulations of moving contact lines, the slip model and the precursor film model each have distinct advantages and disadvantages. Compared to the precursor film model, the slip model is more straightforward and allows for direct extraction of the contact line position and velocity~\cite{bonn2009,xia2020,mhatre2024}. However, it introduces moving boundaries into the computation, which requires the Arbitrary Lagrangian-Eulerian method or other coordinate transformation techniques~\cite{peschka2015,savva2009}. The precursor film model does not incorporate any moving boundary associated with the contact line and easily handles topological changes, such as drop breakup or coalescence~\cite{engelnkemper2016,mhatre2024}.

The microscopic contact line models inherently introduce a multiscale nature to the moving contact line problems, wherein relevant length scales can span approximately six orders of magnitude, ranging from the macroscopic capillary length (millimeter scale) down to the molecular level (nanometer scale)~\cite{snoeijer2013}. Incorporating realistic microscopic lengths into numerical simulations remains challenging due to the prohibitive computational cost. Previously, full-scale simulations with physical microscopic scales were mostly limited to two-dimensional (2D) or axisymmetric situations, typically for steady problems based on Navier-Stokes equations \cite{lowndes1980,sprittles2012,sprittles2015,vandre2012} or unsteady flows using reduced models such as the lubrication equation and the boundary integral equation \cite{snoeijer2010,wang2025,ming2023}. The Navier-Stokes simulations of full-scale unsteady flows with moving contact lines are too expensive to be performed, even for 2D configurations \cite{sui2013validation}. For time-dependent and especially three-dimensional (3D) problems, researchers commonly use artificial slip lengths or precursor film thicknesses that are orders of magnitude larger than actual microscopic values~\cite{diez2002,xia2020,sui2014}. Although such approximations can capture qualitative behaviors, they often lead to quantitative discrepancies compared to reality.

Early works have demonstrated that in slow wetting processes, the fluid interface can generally be divided into three regions~\cite{Voinov1976,hocking1982,hocking1983,cox1986,eggers2005a,eggers2005b,ren2015,zhang2019,Luo2025}. In the outer region, the interface shape is mainly controlled by body forces and flow geometry. In the inner region, the local flow and the interface deformation are significantly influenced by microscopic effects (e.g., slip or precursor film). In the intermediate region, the interface is governed by the balance between viscous and capillary forces, and is hardly affected by microscopic contact line effects and macroscopic configurations. The asymptotic behavior of the interface in the intermediate region, also known as mesoscopic behavior, has been discussed in many studies \cite{hocking1983,eggers2005a,eggers2005b,Luo2025}. In the framework of hydrodynamic theory, the cube of the slope angle in the intermediate region varies logarithmically with respect to the distance to the contact line, known as the Cox-Voinov law \cite{Voinov1976,cox1986}.

In practical applications, the primary focus of interfacial flows is typically on macroscopic behaviors. The computational cost can be significantly reduced by resolving only the macroscopic dynamics and incorporating the small-scale details near the contact line into an appropriate model~\cite{bonn2009,sui2014}. In light of this compromise, various macroscopic methods have been proposed to simulate dynamic wetting processes by incorporating physically realistic microscopic slip lengths. The core of these methods lies in establishing the relationship between dynamic contact angle and contact line velocity. Glasner~\cite{glasner2005} used the Tanner law to solve the contact-line profile, from which the macroscopic drop shape follows under the quasi-static assumption. Building on Glasner's work~\cite{glasner2005}, Savva et al.~\cite{savva2019} derived and used a more accurate Cox law that can distinguish different slip lengths. Somalinga and Bose~\cite{somalinga2000} employed Cox's theory~\cite{cox1986} as a boundary condition to simulate slow wetting flows. A numerical model incorporating the dynamic contact angle was proposed by Afkhami et al.\cite{afkhami2009}, achieving grid-convergent simulations of moving contact lines. Dupont and Legendre\cite{dupont2010} applied the dynamic contact angle at a micrometer scale to model drop spreading, with results in good agreement with experimental observations. Furthermore, Sui et al.~\cite{sui2013} formulated a more rigorous contact-line model and validated its accuracy through quantitative benchmarks against full-scale computations. Solomenko et al.~\cite{solomenko2017} extended the method in Ref.~\cite{sui2013} to 3D geometries.  Qin et al.~\cite{qin2024} and Zhang and Gao~\cite{zhang2024} adopted a different strategy, where the computational domain is truncated inward from the contact line to the intermediate region, thereby bypassing the inner region that consumes the computational cost. They proposed a mesoscopic contact-line model and impose the mesoscopic behavior derived by Hocking \cite{hocking1982,hocking1983} on the truncated boundary.
The accuracy of these macroscopic methods depends on the validity of the underlying contact-line models.

The methods discussed above are all based on the slip models and cannot handle topological changes of the liquid film. For dynamic wetting processes involving film topological changes, precursor film models are preferable, but numerical simulation with a actual microscopic precursor film thickness remains challenging. Additionally, the lubrication approximation is widely applied to model thin film flows since it greatly simplifies the governing equations via the quasi-parallel flow assumptions~\cite{oron1997,qin2018,qin2024,wang2025}. Compared with the full Navier-Stokes simulations, the lubrication equation involves only the film thickness, substantially reducing the computational cost in 3D simulations. The resulting predictions under the lubrication approximation are generally considered valid for moderate contact angles~\cite{snoeijer2013}. Inspired by the mesoscopic contact-line model proposed in Refs.~\cite{qin2024,zhang2024}, we here propose a mesoscopic precursor film model for slow wetting processes under the lubrication approximation. In this model, a thicker precursor film and a thinner (actual microscopic) one exhibit similar interfacial behaviors in the intermediate region. The small yet finite contact angle is introduced through the disjoining pressure. The core of this model lies in modifying the disjoining pressure to dynamically adjust the contact angle on a partially wetting substrate. By solving the lubrication equation with a thicker precursor film, this model significantly reduces the computational cost while ensuring that the macroscopic flow properties remain essentially identical to those obtained with a much thinner precursor film. Moreover, the present model retains the capability to handle topological changes, such as drop coalescence and breakup, offering an advantage over the slip models.

The remainder of this paper is organized as follows. Section~\ref{sec:formu} introduces the mathematical formulation of the lubrication equation and the proposed mesoscopic precursor film model. Section~\ref{sec:nume} details the numerical method. In Section~\ref{sec:exam}, the validity of this method is demonstrated through several numerical examples, accompanied by quantitative comparisons with exact solutions and asymptotic theories. Finally, conclusions are offered in Section~\ref{sec:conc}.

\section{Formulation and mesoscopic precursor film model}
\label{sec:formu}

We consider a thin-film flow on the substrate denoted by the $(x,y)$-plane, where the lubrication approximation is adopted~\cite{oron1997}. The wall is assumed to be stationary. The interface is described by $z=h(\mathbf{x},t)$, where $\mathbf{x}$ denotes the coordinate $(x,y)$.

The contact-line stress singularity is regularized by the precursor film model, which assumes the existence of an ultra-thin liquid film surrounding the macroscopic film~\cite{de1985}. The physical equilibrium precursor film thickness $\bar h_e$ is of molecular scale and remains much smaller than the macroscale characteristic film thickness. The microscopic physics associated with the contact line is incorporated into the governing equations through the disjoining pressure $\Pi$, which depends on the local film thickness $h(\mathbf{x},t)$. Researchers have proposed various forms of $\Pi$. A widely adopted expression of $\Pi$ combines intermolecular attractive and repulsive forces, taking the form:
\begin{equation}
\Pi=-\frac{A}{h^m}+\frac{B}{h^{n}},
\label{dis}
\end{equation}
where 
$m$ and $n$ are the positive integers satisfying $3\leq m<n$~\cite{de1985,oron1997,diez2002,galvagno2014,xia2020}. According to Ref.~\cite{eggers2005a}, the coefficients $A$ and $B$ are positive and related to $\bar h_e$ and the equilibrium contact angle $\bar\theta_e$ as
\begin{equation}
A=\frac{(m-1)(n-1)}{2(n-m)}\sigma\bar h_e^{m-1}\bar \theta_e^2,\quad
B=\frac{(m-1)(n-1)}{2(n-m)}\sigma\bar h_e^{n-1}\bar \theta_e^2,
  \label{AB}
\end{equation}
where $\sigma$ denotes the surface tension. The disjoining pressure can thus be written as
\begin{equation}
    \Pi(h;\bar h_e,\bar\theta_e)=\frac{(m-1)(n-1)}{2(n-m)}\left(-\frac{\bar h_e^{m-1}}{h^m}+\frac{\bar h_e^{n-1}}{h^{n}}\right)\sigma\bar \theta_e^2.
    \label{eq:disj}
\end{equation}
Specific values of $m$ and $n$ exert a negligible quantitative influence on the macroscopic characteristics of the problem~\cite{pismen2008}.
A conventional contact line vanishes when the precursor film model is adopted. In this work, the position of the effective contact line is defined as the location where the film interface exhibits its local maximum curvature~\cite{xia2020}.

The evolution of $h(\mathbf{x},t)$ is governed by the lubrication equation~\cite{oron1997,eggers2005a}
\begin{equation}
\frac{\partial h}{\partial t}+\nabla \cdot \left(h~\mathbf{u}_{avg}\right)
=0,\quad \mathbf{u}_{avg}=\frac{\sigma}{3\mu}h^2\nabla\left[\nabla^2 h-\phi+\Pi(h;\bar h_e,\bar \theta_e)\right],
  \label{lub1}
\end{equation}
where $\mu$ denotes the liquid viscosity, and $\phi$ is the body-force potential. In numerical simulations, the computational domain is extended to the far-field region encompassing the equilibrium precursor film where the film profile remains uniform and flat, yielding the following boundary conditions:
\begin{equation}
 \mathbf{n}_b \cdot\nabla h = 0, \quad  \mathbf{n}_b\cdot \nabla (\nabla^2 h)  = 0,
\label{bc1}
\end{equation}
where $\mathbf{n}_b$ is the unit normal vector pointing outward from the border.

Numerically solving problem~\eqref{lub1} and \eqref{bc1} is generally computationally expensive. Because of the extremely small precursor film thickness $\bar h_e$, a highly refined mesh is required near the  contact line to capture its dynamics. Motivated by Refs.~\cite{qin2024,zhang2024}, we propose a mesoscopic precursor film model, wherein a thicker precursor film exhibits similar interfacial behaviors in the intermediate region to those of a thinner precursor film. By employing a thicker precursor film thickness $h_e$, the proposed model aims to reduce the heavy computational cost while ensuring that the macroscopic flow properties remain essentially identical to those obtained with a thinner precursor film thickness $\bar h_e$. A sketch of the mesoscopic precursor film model is presented in Fig.~\ref{Fig1}. The detailed implementation procedure of this model is shown below.

\begin{figure}
  \centering
  \includegraphics[scale=0.6,trim=0cm 3cm 0cm 0cm,clip]{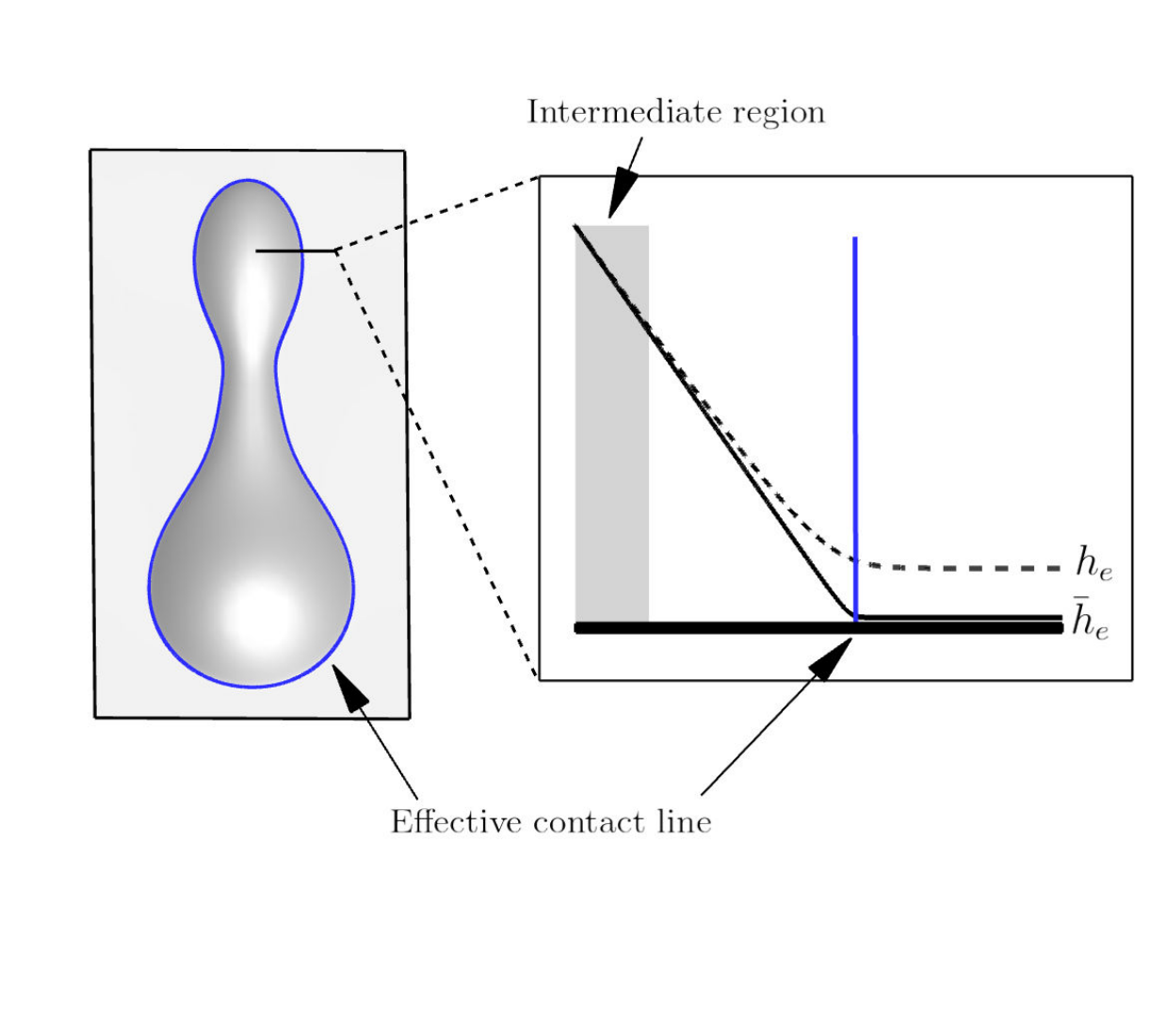}
  \caption{
  Schematic of the mesoscopic precursor film model: the interface profile with a thicker precursor film ($h_e$) exhibits similar behavior within the intermediate region to that of the physically realistic, thinner film ($\bar{h}_e$).
}
  \label{Fig1}
\end{figure}

Within the precursor film model, the asymptotic behavior of the interface in the intermediate region is independent of the macroscopic effects and have been discussed in early studies~\cite{eggers2005a,pismen2008}, i.e.,
\begin{equation}
|\nabla h|^3\approx\bar\theta_e^3+\frac{9\mu}{\sigma}u_{cl}\ln\frac{\alpha x^*\bar\theta_e }{\bar h_e }, \label{22}
\end{equation}
where $x^*$ is the distance from the contact line, $\alpha $ is a constant depending on $m$ and $n$, and $u_{cl}$ is the  contact line velocity relative to the substrate. We stipulate that $u_{cl}>0$ for advancing contact lines and $u_{cl}<0$ for receding contact lines.

We next consider the film dynamics with an artificially thicker precursor film characterized by a uniform thickness $h_e$ and a dynamic contact angle $\theta_d$ that is locally adjustable. The film evolution is still governed by the lubrication equation
\begin{equation}
    \frac{\partial h}{\partial t}+\nabla \cdot \left(h~\mathbf{u}_{avg}\right)
=0,\quad \mathbf{u}_{avg}=\frac{\sigma}{3\mu}h^2\nabla\left[\nabla^2 h-\phi+\Pi(h;h_e,\theta_d)\right],\label{lub22}
\end{equation}
where the disjoining pressure is still formally given by Eq.~\eqref{eq:disj} but parameterized by $h_e$ and $\theta_d$. While $h_e$ is constant, the dynamic contact angle $\theta_d$ can vary with time and space and should be carefully selected. It is expected that the disjoining pressure primarily acts near the contact line, where we prefer that $\theta_d$ does not depend on the distance to the contact line such that the local dynamics can still be described by the one-dimensional theory of Eggers \cite{eggers2005a}. Specifically, in the intermediate region near the contact line, we have
\begin{equation}
    |\nabla h|^3\approx\theta_d^3+\frac{9\mu}{\sigma}u_{cl}\ln\frac{\alpha x^* \theta_d }{ h_e }. \label{22222}
\end{equation}
To ensure that the contact line velocities  and the asymptotic behaviors in the intermediate region for thin and thick precursor films are consistent, we match the asymptotic behaviors \eqref{22} and \eqref{22222} and derive the relationship
\begin{equation}
\theta_d^3 = \bar\theta_e^3 + \frac{9\mu}{\sigma} u_{cl} \ln \left( \frac{{h}_e}{\bar h_e} \frac{\bar{\theta}_e}{\theta_d} \right),
\label{eq:mapping}
\end{equation}
which can be used to determine $\theta_d$ near the contact line provided that the remaining parameters are known. 

A direct way to determine $u_{cl}$ is to track the contact line, i.e., $u_{cl}=\mathbf{n}_{cl}\cdot d\mathbf{x}_{cl}/dt$, where $\mathbf{x}_{cl}$ is the contact line position and $\mathbf{n}_{cl}$ is the outward normal to the contact line. This procedure, however, requires an explicit extraction of the contact line, which is not convenient under the precursor film model. Instead, $u_{cl}$ is globally extended to
\begin{equation}
  u_{cl} = \frac{h}{h-h_e}\mathbf{u}_{avg}\cdot\mathbf{n},
  \label{eq:ucl}
\end{equation}
where $\mathbf{n}$ is the vector field directing to $-\nabla h$. Near the contact line, we require $\mathbf{n}=\mathbf{n}_{cl}$  and can easily verify that Eq.~\eqref{eq:ucl} is the one-dimensional equation governing the local film profile, equivalent to Eq.~(7) of Ref.~\cite{eggers2005a}. An additional advantage of Eq.~\eqref{eq:ucl} is that it naturally extends $u_{cl}$ (and thus $\theta_d$) into a global field function, facilitating the numerical solution of Eq.~\eqref{lub22}. We expect that $u_{cl}$ given by Eq.~\eqref{eq:ucl} converges to the contact-line velocity when the distance to the contact line is sufficiently small, as confirmed numerically in the following. Note that $u_{cl} \neq \mathbf{u}_{avg}\cdot\mathbf{n}_{cl}$ at the contact line due to the mass flux carried by the precursor film. Away from the contact line, $u_{cl}$ calculated by Eq.~\eqref{eq:ucl} does not play a role in the solution since the disjoining pressure can be neglected.

Denoting $R$ as the characteristic lateral length of the macroscopic film, we introduce the following nondimensionalization scheme,
\begin{equation}
(X,Y)=\frac{1}{R}(x,y), \quad (H,H_e,\bar H_e)=\frac{1}{\bar \theta_eR}(h,h_e,\bar h_e), \quad T=\frac{\sigma \bar \theta_e^3}{3\mu R}t, \quad \Phi=\frac{R}{\bar\theta_e}\phi,\quad \Theta=\frac{\theta_d}{\bar\theta_e}.
  \label{non}
\end{equation}
Then the lubrication equation \eqref{lub22} with the thicker precursor film can be written as
\begin{subequations}
\begin{align}
&\frac{\partial H}{\partial T}+\nabla \cdot \left(H\mathbf{U}_{avg}\right) = 0,  \label{444}\\
&\mathbf{U}_{avg}=H^2 \nabla \left[\nabla^2 H-\Phi+\frac{(m-1)(n-1)}{2(n-m)}\left(-\frac{H_e^{m-1}}{H^m}+\frac{H_e^{n-1}}{H^n}\right)\Theta^2\right],
  \label{4}
\end{align}
\end{subequations}
supplemented by 
\begin{equation}
\Theta^3 = 1 + 3U_{cl}\ln\frac{K}{\Theta}, \quad \text{with} \quad U_{cl}=\frac{H}{H-H_e}\mathbf{U}_{avg}\cdot\mathbf{n}.
  \label{5}
\end{equation}
Here $K=h_e/\bar h_e=H_e/\bar H_e$ is the ratio of the precursor film thicknesses, and is prescribed as a fixed value (e.g., $100$) in numerical simulations. For $K=1$, we have $\Theta=1$ according to Eq.~\eqref{5}, and the proposed mesoscopic precursor film model reduces to the original model.

In summary, by modifying the disjoining pressure with a dynamic contact angle, the intermediate and hence macroscopic interfacial behavior of a physically realistic thin precursor film can be modeled using a much thicker one.

\section{Numerical method}
\label{sec:nume}

In this work, we follow Refs.~\cite{eggers2005a,pismen2008} and choose $m = 3,n=5$ under which we have $\alpha=\text{e}/4$ in Eqs.~\eqref{22} and \eqref{22222} such that theoretical validation becomes more convenient. Notably, the present framework applies to more general configurations of $m$ and $n$.

The core of the proposed mesoscopic precursor film model relies on an appropriate disjoining pressure modification factor, i.e., $\Theta$. By performing numerical simulations of the spreading process of axisymmetric drops, Appendix \ref{appendix1} identifies the precise active region of $\Theta$. The results indicate that $\Theta$ plays a significant role only within a narrow, localized zone near the contact line, specifically for $H \in [1.1, 20] H_e$. Beyond this range, modifying the disjoining pressure exerts a negligible impact on the macroscopic film dynamics. Moreover, the macroscopic film dynamics exhibit high sensitivity to the lower bound of this localized region: even a minor shift in the lower bound from $1.1H_e$ to $1.5H_e$ leads to a noticeable deviation. Consequently, a standard decoupled solution procedure fails to yield accurate results, where $\Theta$ is first calculated explicitly from the relationship \eqref{5} and subsequently substituted into the governing equation \eqref{444} and \eqref{4} for the next time step. During time integration, such an explicit decoupling inevitably causes a spatial misalignment of the $\Theta$-active window, thereby leading to a significant deviation in the macroscopic film dynamics.

Equations \eqref{444}, \eqref{4} and \eqref{5} should be solved in a coupled manner. However, when solving the coupled system of \eqref{444}, \eqref{4} and \eqref{5}, the denominator $(H - H_e)$ in relationship \eqref{5} vanishes in the flat precursor-film region where $H=H_e$, leading to numerical divergence. To regularize this singularity as well as for numerical convenience, given that the contact line velocity $U_{cl}$ is typically small, we approximate Eq.~\eqref{5} to
\begin{equation}
\Theta^2 \approx  1 + \frac{2H}{H-H_e}\mathbf{U}_{avg}\cdot\mathbf{n}\ln\frac{K}{\Theta}.
  \label{approx}
\end{equation}
Substituting \eqref{approx} back into the governing equation \eqref{444} and \eqref{4} yields the regularized system:
\begin{subequations}
\begin{align}
&\frac{\partial H}{\partial T}+\nabla \cdot \left(H\mathbf{U}_{avg}\right) = 0, \label{111}\\
&\mathbf{U}_{avg} = H^2\nabla\left[\nabla^2 H-\Phi-\frac{2 H_e^2}{H^3}+\frac{2 H_e^4}{H^5}-\frac{4 H_e^2(H+ H_e)}{H^4}\mathbf{U}_{avg}\cdot\mathbf{n}\ln\frac{K}{\Theta}\right], \label{222}
\end{align}
\end{subequations}
wherein the vanishing denominator $(H - H_e)$ is analytically eliminated except in the logarithmic function. Note that this approximation does not violate the cubic dependence of the film slope on the contact line velocity in the intermediate region, i.e., Eqs.~\eqref{22} and \eqref{22222}.

We employ a finite element method to solve the coupled problem \eqref{111} and \eqref{222}. By introducing an auxiliary pressure $f$ with $\mathbf{U}_{avg} = H^2\nabla f$ to decompose the fourth-order equation into two coupled second-order equations and incorporating the natural boundary conditions derived from \eqref{bc1}, the weak forms of \eqref{111} and \eqref{222} are expressed as:
\begin{subequations}
\begin{align}
&\int_\Omega \frac{\partial H}{\partial T}v_1{\rm d}\Omega-\int_\Omega H^3\nabla f\cdot \nabla v_1 {\rm d}\Omega
=0,\label{1111}\\
&\int_\Omega \left[
f+\Phi+\frac{2 H_e^2}{H^3}-\frac{2 H_e^4}{H^5}+\frac{4 H_e^2(H+ H_e)}{H^2}\nabla f\cdot\mathbf{n}\ln\frac{K}{\Theta}
\right]v_2{\rm d}\Omega+\int_\Omega \nabla H\cdot \nabla v_2 {\rm d}\Omega=0,\label{2222}
\end{align}
\end{subequations}
where $\Omega$ is the computational domain, and $v_1,v_2$ represent test functions.

\begin{figure}
  \centering
  \includegraphics[scale=0.4]{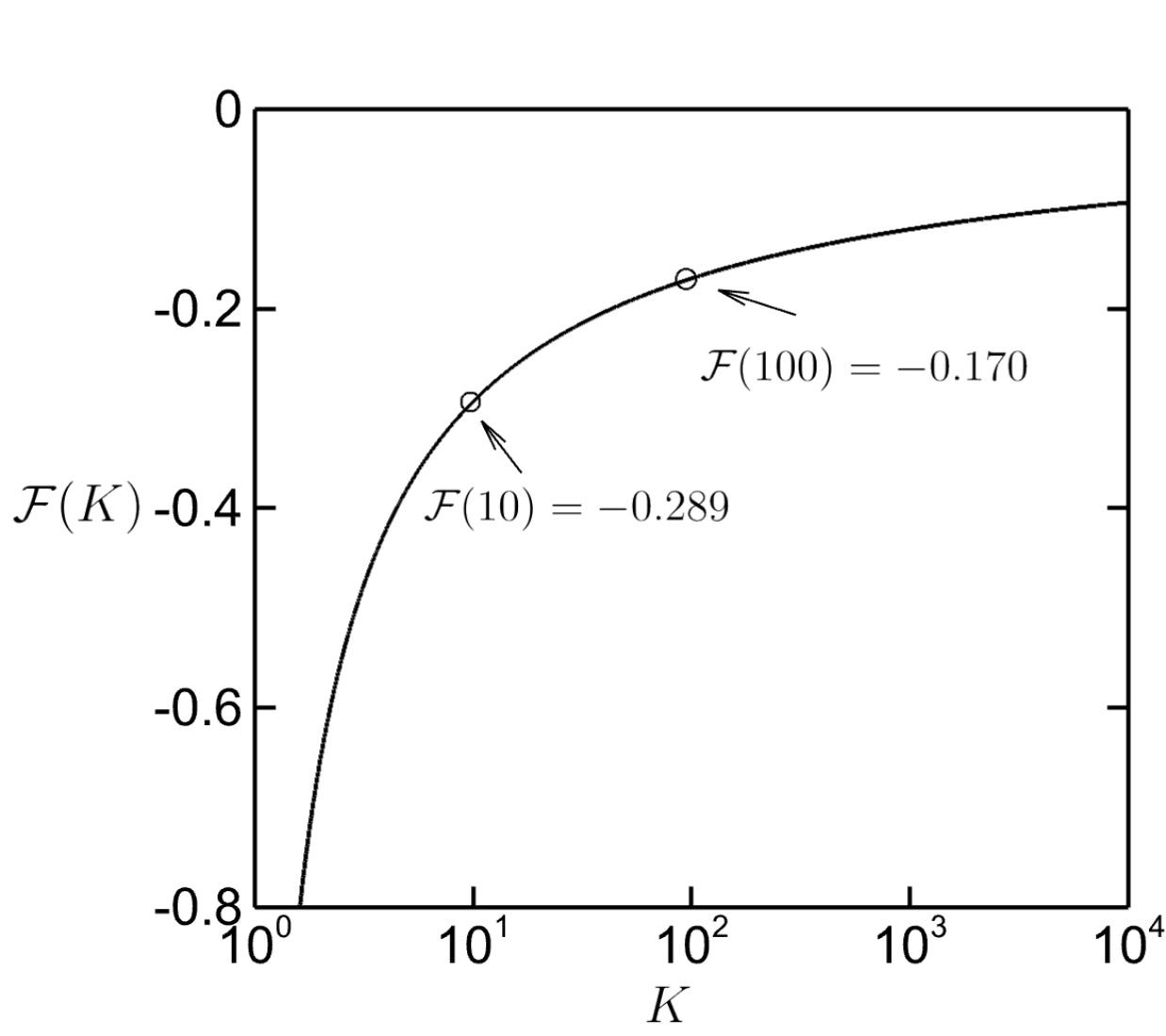}
  \caption{
  Variation of $\mathcal F(K)$ against the scaling ratio $K$ (on a logarithmic scale).
}
  \label{Fig3}
\end{figure}

Regarding the remaining variables $\mathbf{n}$ and $\Theta$ in \eqref{2222}, a natural approach is to treat them explicitly. To define a global vector filed $\mathbf{n}$ throughout the entire domain, we utilize the normalized negative gradient of $H$:
\begin{equation}
\mathbf{n}=\frac{-\nabla H}{\sqrt{|\nabla H|^2+\epsilon}},
\label{NxNy}
\end{equation}
where $\epsilon$ is a small regularization parameter introduced to prevent division by zero in the flat precursor-film region. This definition naturally reduces to $\mathbf{n}_{cl}$ in the vicinity of the  contact line.

When explicitly evaluating $\Theta$ via relationship \eqref{5}, two numerical issues arise. First, the denominator $(H - H_e)$ vanishes in the flat precursor-film region. Second, for $\Theta>0$, the algebraic term $(\Theta^3-1)/\ln (K/\Theta)$ possesses a minimum value denoted as $\mathcal{F}(K)$. The variation of $\mathcal{F}(K)$ as a function of $K$ is illustrated in Fig. \ref{Fig3}. Consequently, relationship \eqref{5} becomes unsolvable at any grid point where $3U_{cl}$ drops below the threshold $\mathcal{F}(K)$. Physically, this indicates that the velocity of a receding contact line cannot exceed a threshold. To circumvent these two issues, we regularize and modify the relation \eqref{5} to
\begin{equation}
\Theta^3 = 1 + {\rm max}\left[
\frac{3H^3}{\sqrt{(H - {H}_e)^2 + {H}_e^2 / 10}} \nabla f \cdot \mathbf{n}
, \mathcal{F}(K)
\right]
\ln\frac{K}{\Theta}.
  \label{55}
\end{equation}
In our simulations, the regularization parameters ${H}_e^2 / 10$ and $\epsilon=10^{-3}$ are verified to ensure numerical stability without sacrificing accuracy.

In summary, an explicit-implicit (or semi-implicit) time-stepping scheme is adopted: the remaining $\mathbf{n}$ and $\Theta$ in \eqref{222} are treated explicitly using values from the current time step $T^n$, while the film thickness $H$ and auxiliary pressure $f$ are solved implicitly at the next time step $T^{n+1}$. The temporal derivative of the film thickness is approximated via the first-order backward difference:
\begin{equation}
\frac{\partial H}{\partial T}(\mathbf{X}, T^{n+1}) \approx \frac{H(\mathbf{X}, T^{n+1}) - H(\mathbf{X}, T^n)}{\Delta T}.
\end{equation}

The detailed solution procedure is summarized as follows:

\begin{itemize}
    \item[] \textbf{Initialization and relaxation:}

    \begin{itemize}
    \item[1.] Generate initial domain $\Omega^{0}$ and mesh $\mathbf{X}^{0}$. Initialize $H^{init}$ and $f^{init}$.
    \item[2.] Relax the system for a few small time steps without applying the mesoscopic precursor film model, allowing the initial non-physical $H^{init}$ and $f^{init}$ to relax into a physical state ($H^{0}$ and $f^{0}$).
    \item[3.] Initialize $T^0 = 0,~\mathbf{n}^0=\mathbf{0},~\Theta^0=1$.
    \end{itemize}

    \item[] \textbf{Do} $n = 0,1,2,\dots$

    \begin{itemize}
    \item[1.] Select the time step $\Delta T$ adaptively based on the initial Newton residual of the previous step: $T^{n+1} = T^n + \Delta T$.
    \item[2.] With explicitly prescribed $\mathbf{n}(T^{n})$ and $\Theta(T^{n})$, solve the coupled weak forms \eqref{1111} and \eqref{2222} for $H(T^{n+1})$ and $f(T^{n+1})$ using Newton's iterative method.
    \item[3.] Update the vector field $\mathbf{n}(T^{n+1})$ from
     $H(T^{n+1})$ via \eqref{NxNy}. Then solve the regularized relationship \eqref{55} for $\Theta(T^{n+1})$ using $H(T^{n+1}),~f(T^{n+1}),~\mathbf{n}(T^{n+1})$.
    \item[4.] Linearly stretch or shrink the mesh according to the position of the  contact line at each time step. Additionally, perform adaptive remeshing based on the distribution of $H^{-1}$ every 20 time steps.
    \end{itemize}

    \item[] \textbf{End Do}
\end{itemize}

The algorithm is implemented in FreeFem++~\cite{hecht2012}, an open-source finite element solver well-suited for moving-mesh problems~\cite{qin2024,decoene2012}. The mesh moving is realized via built-in function \texttt{movemesh}, and the adaptive remeshing is executed by the built-in function \texttt{adaptmesh}. This remeshing is driven by the spatial distribution of $H^{-1}$, which automatically refines the mesh in the vicinity of the  contact line.

\section{Examples}
\label{sec:exam}

The present method is validated by examining the spreading, retraction, sliding, and coalescence of the drops, as well as the breakup of a liquid ridge. Both 2D and 3D configurations are considered. For comparison, the precursor film thickness is varied among $3\times10^{-3}$, $2\times10^{-3}$, $10^{-3}$, $10^{-4}$ and $10^{-5}$, where $10^{-5}$ is considered as the same order of magnitude as the physically realistic value for millimeter-sized drops or liquid ridges. For convenience, the mesoscopic precursor film model is referred to as the "Meso-Model" (all targeting $\bar H_e=10^{-5}$ with different values of $K$), while the original model is designated as the "Orig-Model" ($K=1$ with different values of $\bar H_e$).

All numerical test cases have undergone time-step and mesh-independence validations. The results demonstrate that a minimum mesh size equal to the precursor film thickness $H_e$ is sufficient to ensure numerical accuracy.

\subsection{Axisymmetric drop spreading and retraction}
\label{sec:asym}

The first example considers the spreading or retraction of the axisymmetric drops, which can be reduced to a 1D equation characterized by a single advancing or receding contact line~\cite{jung2003,mistry2015,qin2023}. Formulations of the axisymmetric drop spreading and retraction are detailed in Appendix \ref{appendix2}. The drop exhibits spreading when the initial radius $R_0 < 1$, corresponding to an advancing contact line, whereas it undergoes retraction when $R_0 > 1$, corresponding to a receding contact line.

\subsubsection{Axisymmetric drop spreading}
\label{sec:sprea}

To evaluate the performance of the Meso-Model under an advancing contact line, the initial drop radius is set to $R_0=\sqrt{3}/2 < 1$. Five sets of numerical simulations are performed for comparison: three using the Orig-Models with $\bar H_e=10^{-3}$, $10^{-4}$, and $10^{-5}$, and the other two using the Meso-Models (targeting $\bar H_e=10^{-5}$) with $K=100$ and $K=10$.

\begin{figure}
  \centering
  \includegraphics[scale=0.4]{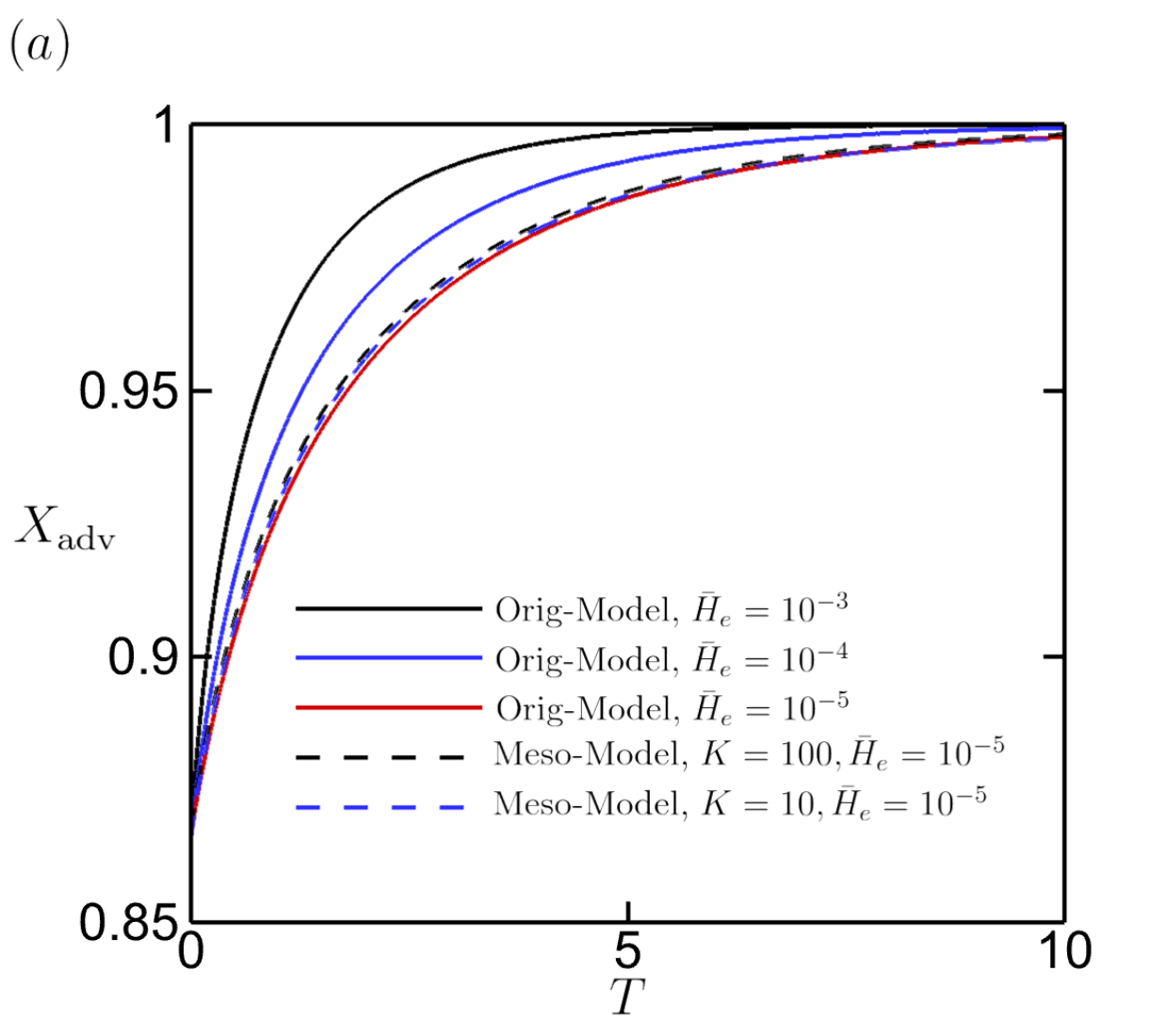}\includegraphics[scale=0.4]{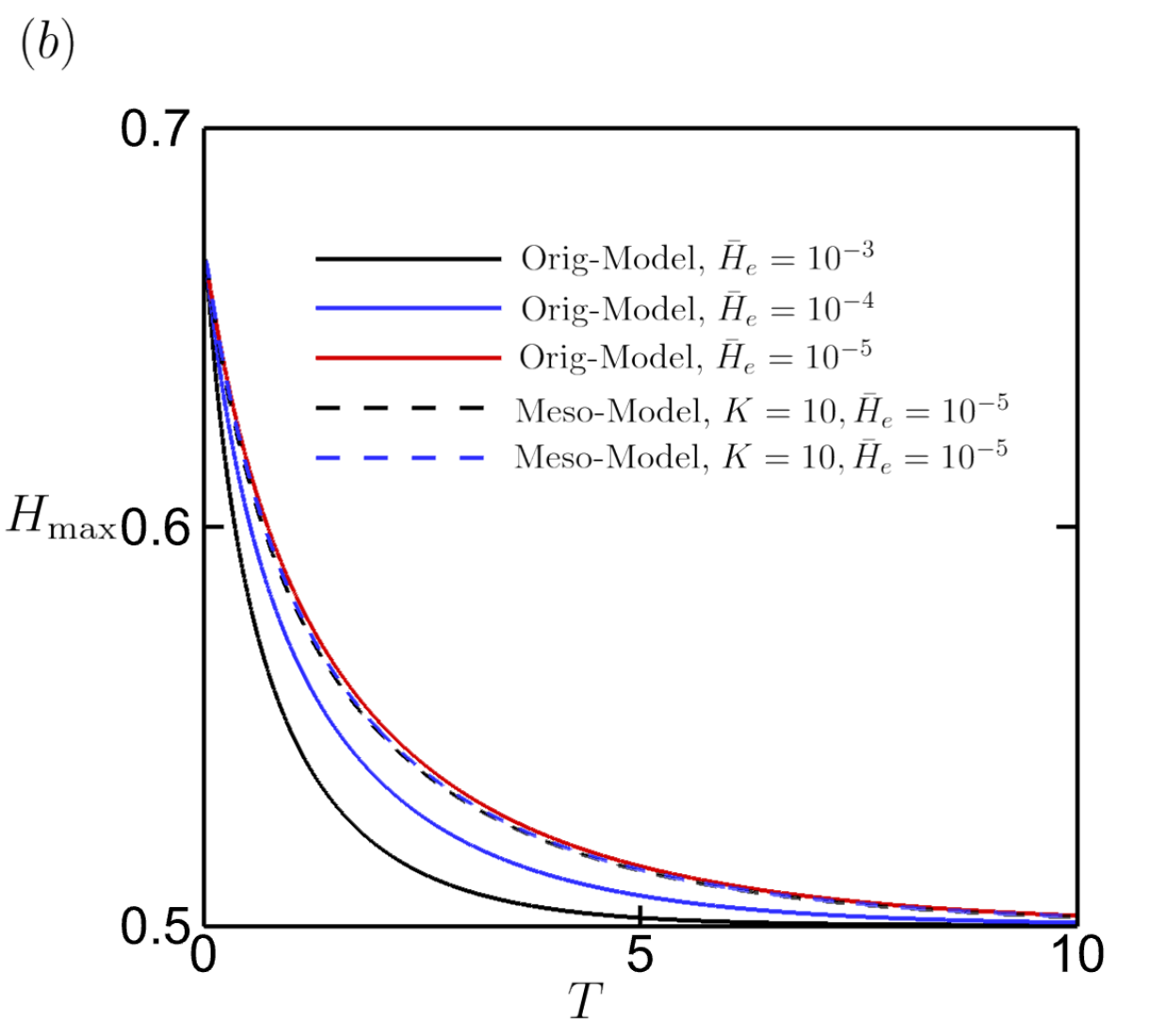}
  \caption{
  Axisymmetric drop spreading: temporal evolution of the (\textit{a}) drop radius (advancing contact line position $X_{adv}$) and (\textit{b}) maximum drop height $H(0)$. The solid and dashed lines denote the Orig-Models and Meso-Models, respectively.
}
  \label{Fig4}
\end{figure}

\begin{figure}
  \centering
  \includegraphics[scale=0.4]{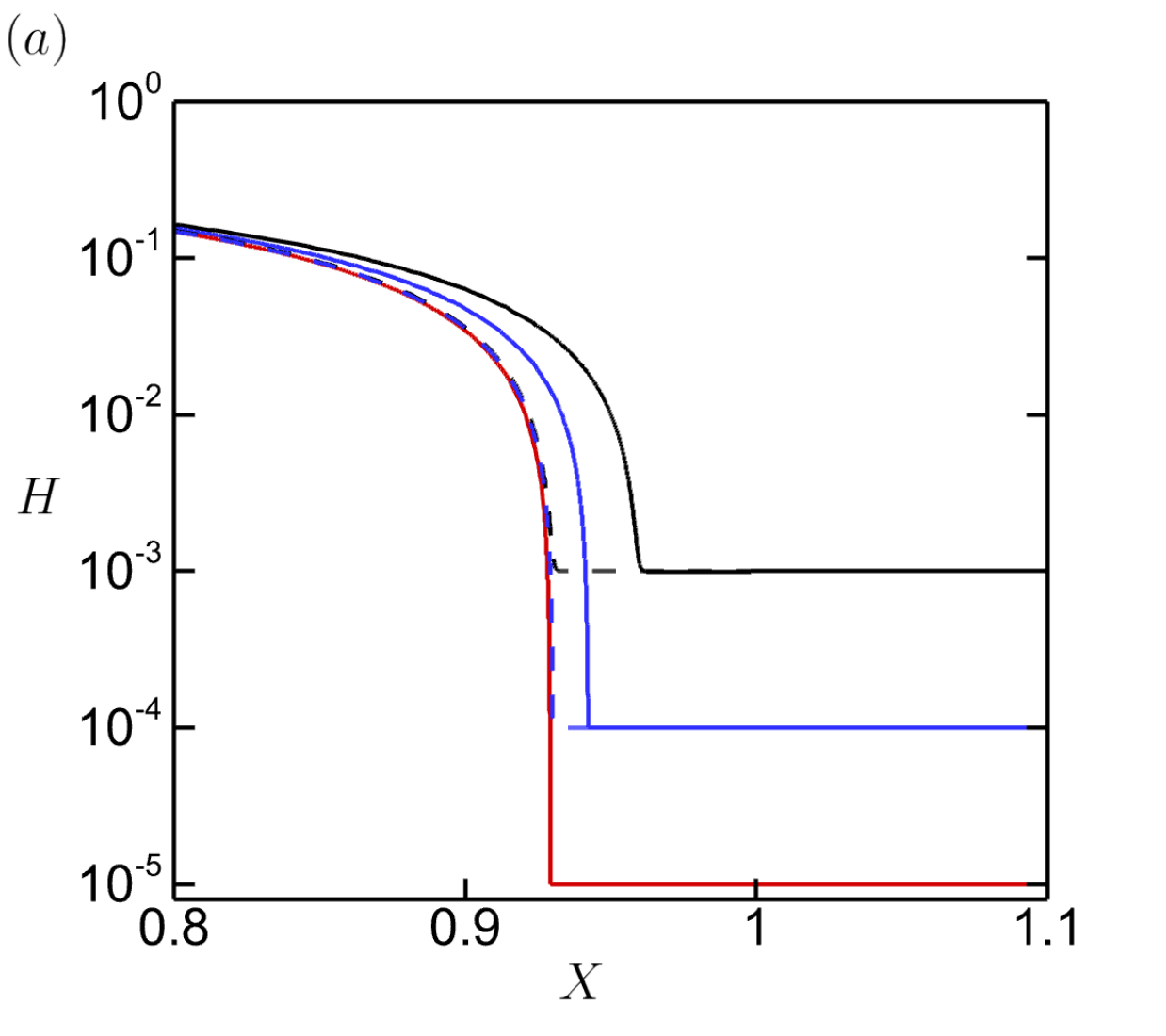}\includegraphics[scale=0.4]{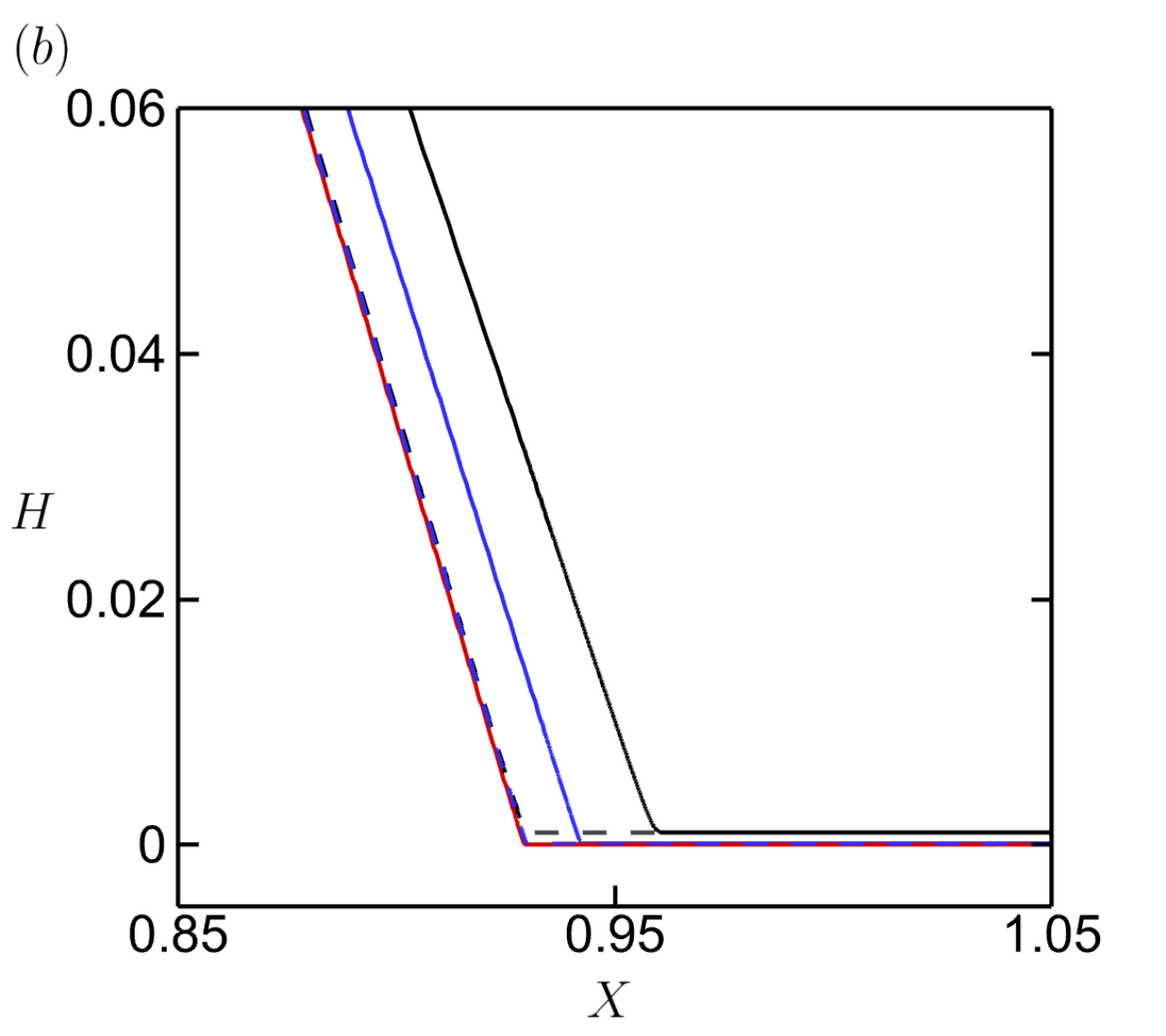}
  \caption{
  Drop interface profiles near the  contact line during axisymmetric drop spreading at $T=1.0$ with (\textit{a}) a logarithmic $H$-axis and (\textit{b}) a linear $H$-axis. The legend is identical to that in Fig.~\ref{Fig4}.
}
  \label{Fig5}
\end{figure}

Fig.~\ref{Fig4} presents the temporal evolution of the macroscopic drop parameters, including the drop radius (equal to the advancing contact line position $X_{adv}$) and the maximum drop height $H(0)$. As illustrated, a smaller $\bar H_e$ results in a slower spreading rate, whereas the Orig-Model with $\bar H_e=10^{-3}$ exhibits the fastest spreading and is considered to deviate most from the physical reality. The macroscopic evolutions of the Meso-Models agree well with those of the corresponding Orig-Models.

\begin{figure}
  \centering
  \includegraphics[scale=0.4]{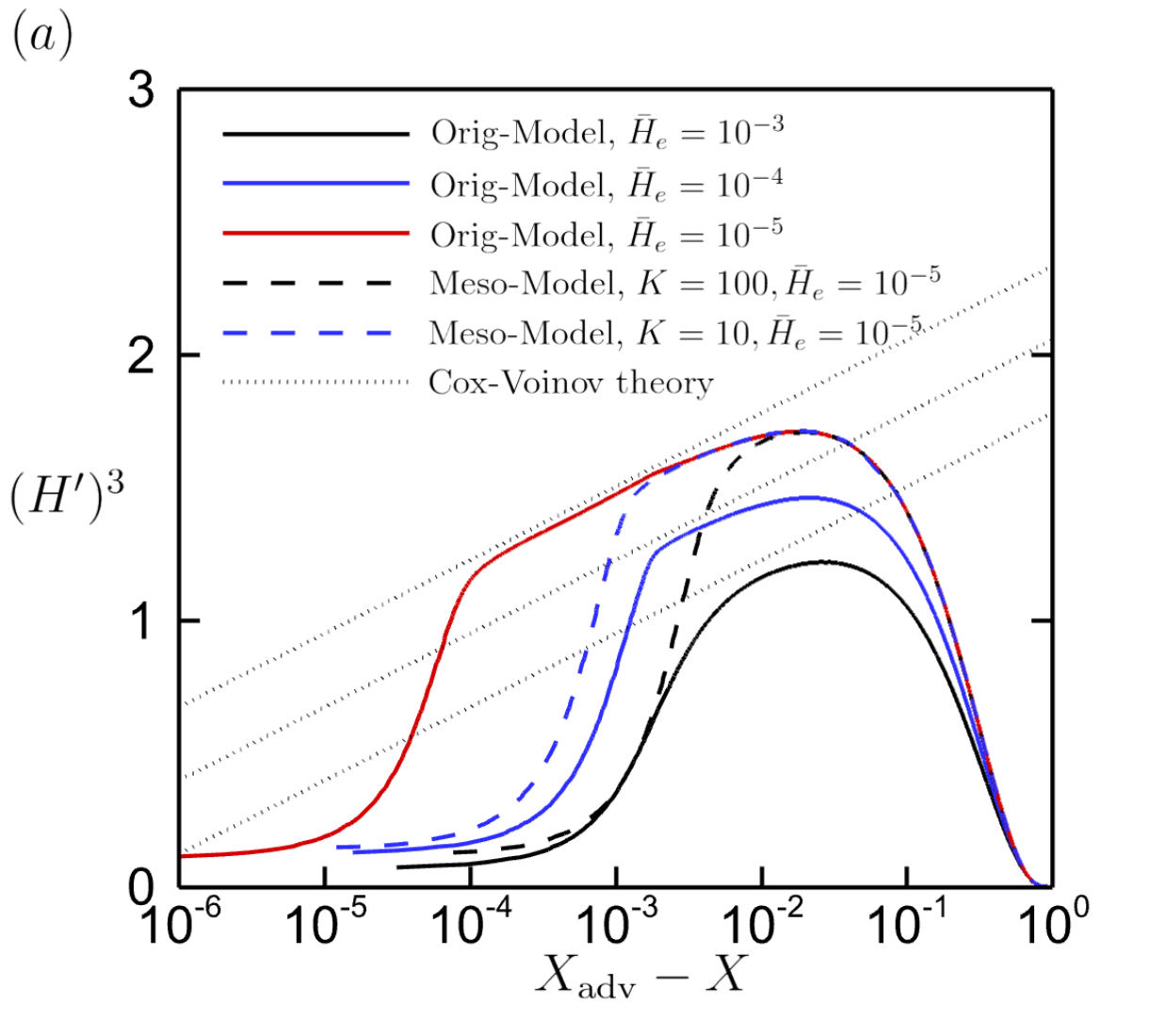}\includegraphics[scale=0.4]{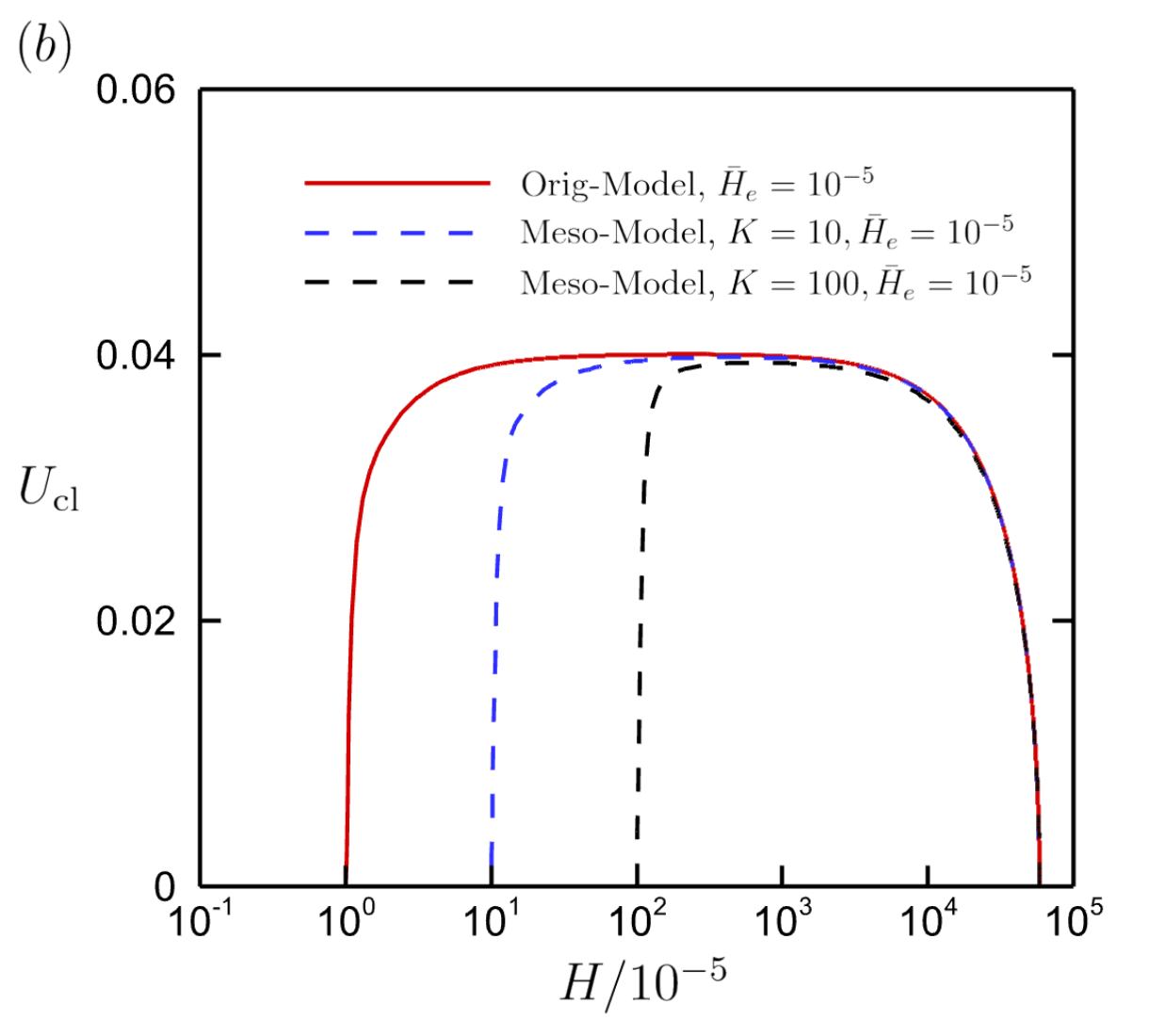}
  \caption{
  Profiles near the  contact line for axisymmetric drop spreading at a velocity of $0.04$: (\textit{a}) distribution of the third power of the interface slope plotted against the distance to the  advancing contact line with a logarithmic $X$-axis; (\textit{b}) extended contact line velocity $U_{cl}$ as a function of the drop thickness $H \times 10^5$ (only the Orig-Model with $\bar H_e = 10^{-5}$ and the Meso-Model with $K = 10$ and $100$ are shown).
}
  \label{Fig6}
\end{figure}

Fig.~\ref{Fig5} shows the drop interface profiles near the  contact line at $T=1.0$, which clearly illustrates that the Meso-Models and the Orig-Model exhibit similar interface behaviors in the intermediate region. To further investigate the interface slope variation described by the Cox--Voinov theory \cite{eggers2005a}
\begin{equation}
(H')^3\approx1+3U_{cl}\ln\frac{\text{e} \left(X_{adv}-X\right)}{4\bar H_e},
\end{equation}
we compare in Fig.~\ref{Fig6}(a) the distribution of $(H')^3$ among the Meso-Models, the Orig-Models and the Cox-Voinov theory at an identical advancing contact line velocity of $0.04$. First, a smaller $\bar H_e$ leads to a more pronounced intermediate region as described by the Cox-Voinov theory. Second, the Meso-Models accurately match the interface behavior of the corresponding Orig-Models within the intermediate region. Furthermore, Fig.~\ref{Fig6}(b) depicts the spatial distribution of the extended contact line velocity $U_{cl}$. Notably, a distinct intermediate platform exists where $U_{cl}$ remains nearly constant. This plateau implies that the disjoining pressure correction factor $\Theta$ is also invariant within this region, further substantiating the validity of the Meso-Models.

\subsubsection{Axisymmetric drop retraction}

To evaluate the performance of the Meso-Model under a receding contact line, the initial drop radius is set to $R_0=1.2>1$, which drives the drop to retract toward its equilibrium state. Correspondingly, five sets of numerical simulations are performed using identical parameter configurations to those in the spreading case in Section \ref{sec:sprea}.

\begin{figure}
  \centering
  \includegraphics[scale=0.4]{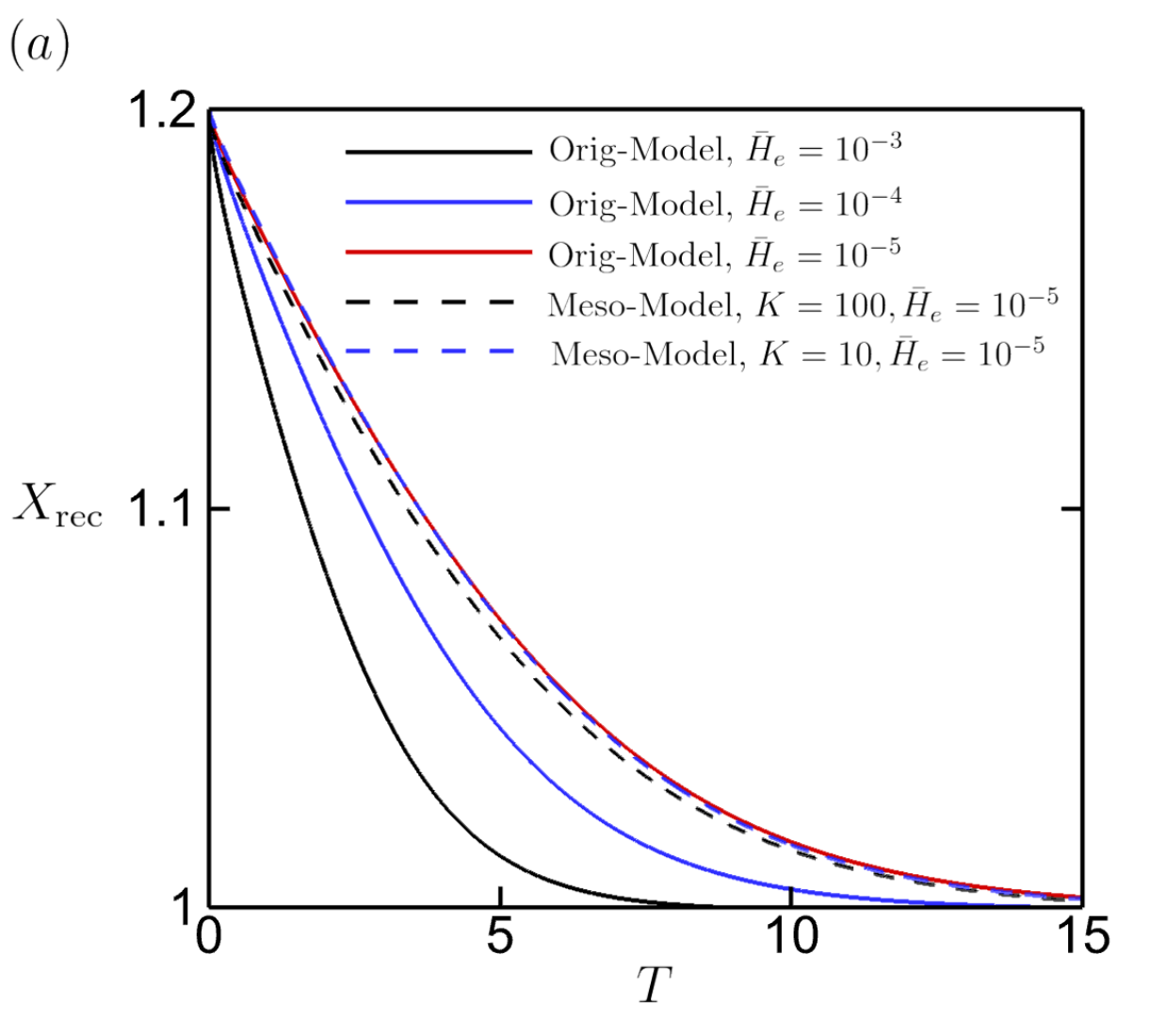}\includegraphics[scale=0.4]{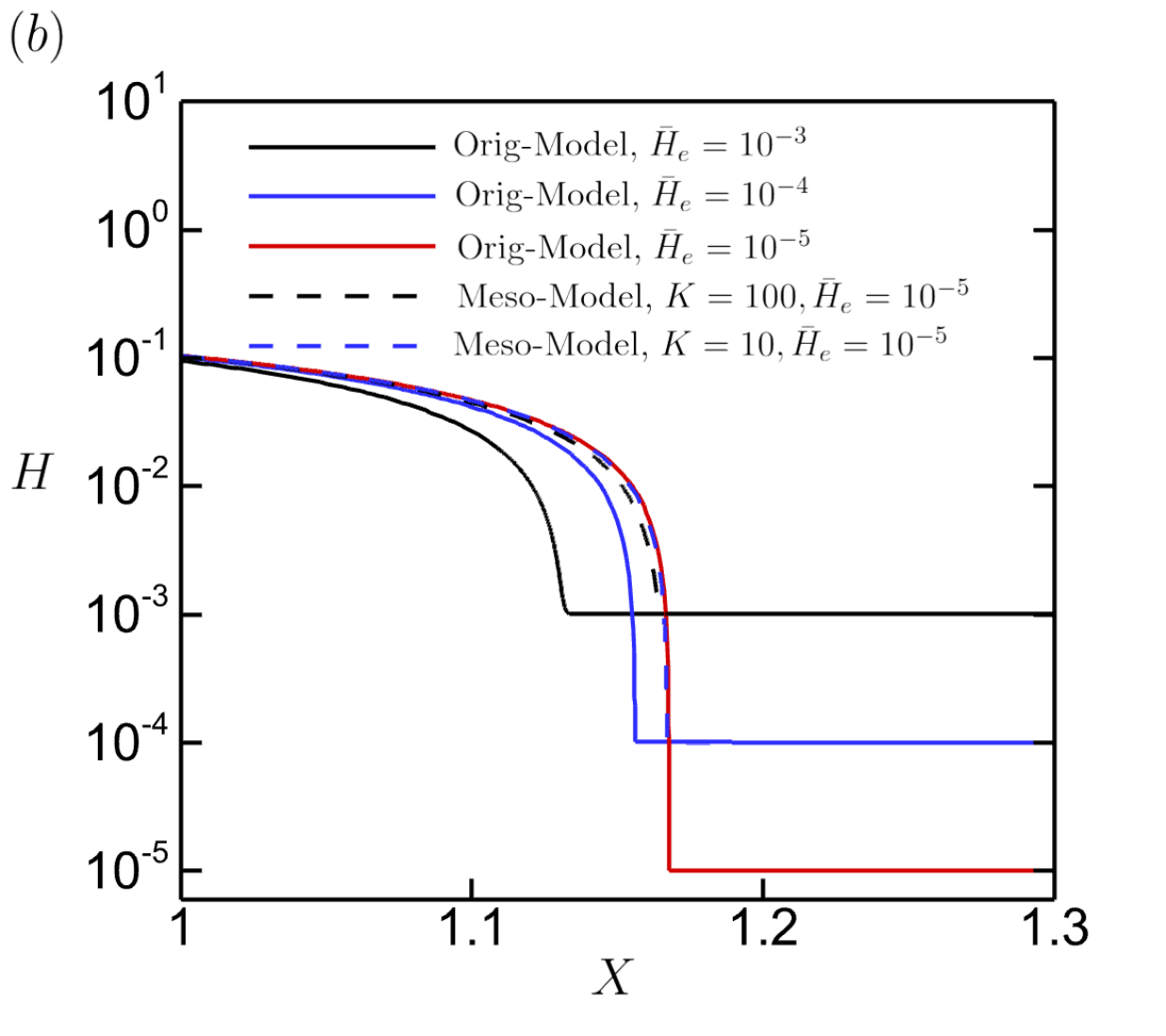}
  \caption{
  Axisymmetric drop retraction: (\textit{a}) temporal evolution of the drop radius (receding contact line position $X_{rec}$); (\textit{b}) drop interface profiles near the  receding contact line during axisymmetric drop retraction at $T=1.0$ with a logarithmic $H$-axis.
  }
  \label{Fig7}
\end{figure}

\begin{figure}
  \centering
  \includegraphics[scale=0.4]{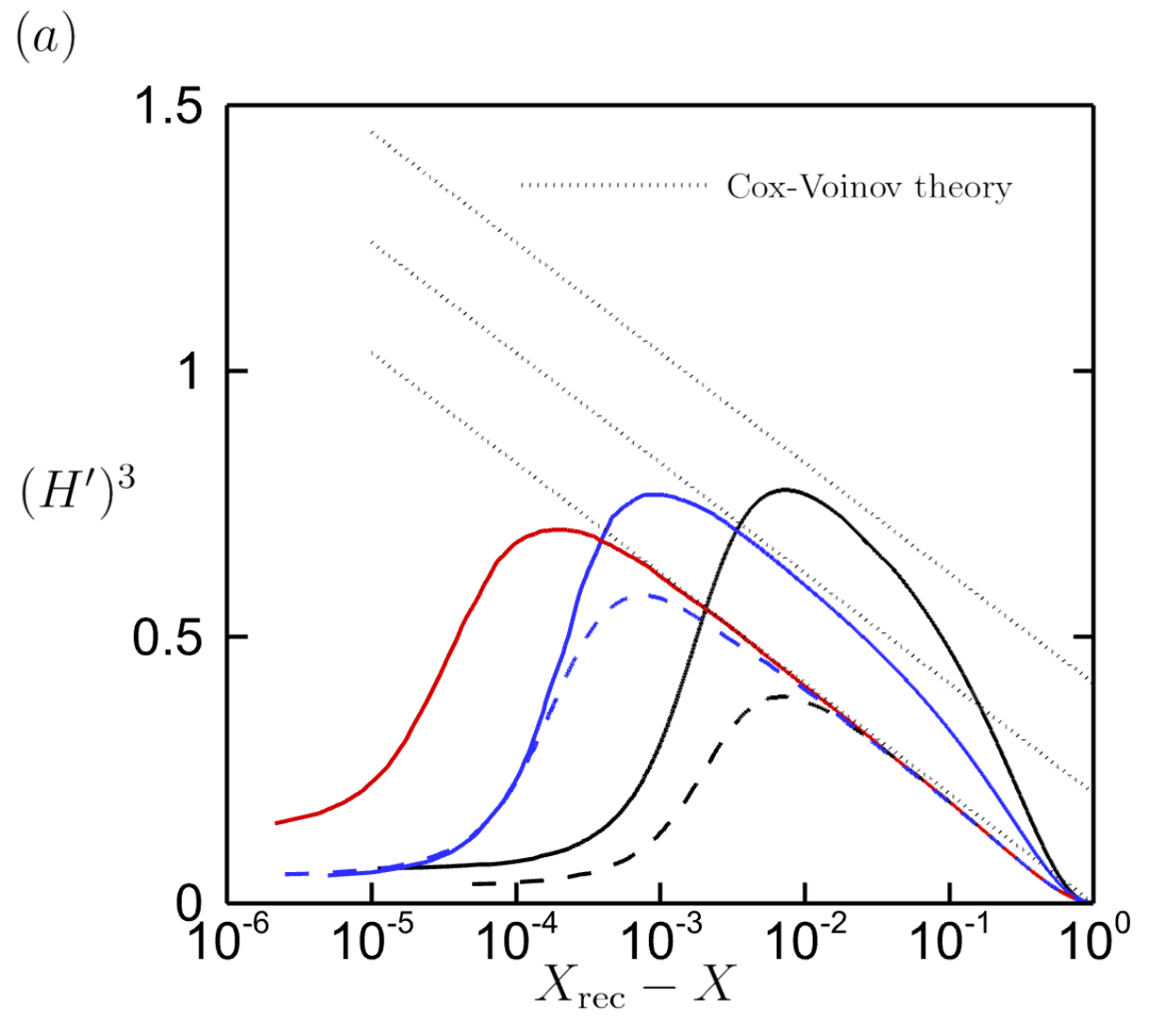}\includegraphics[scale=0.4]{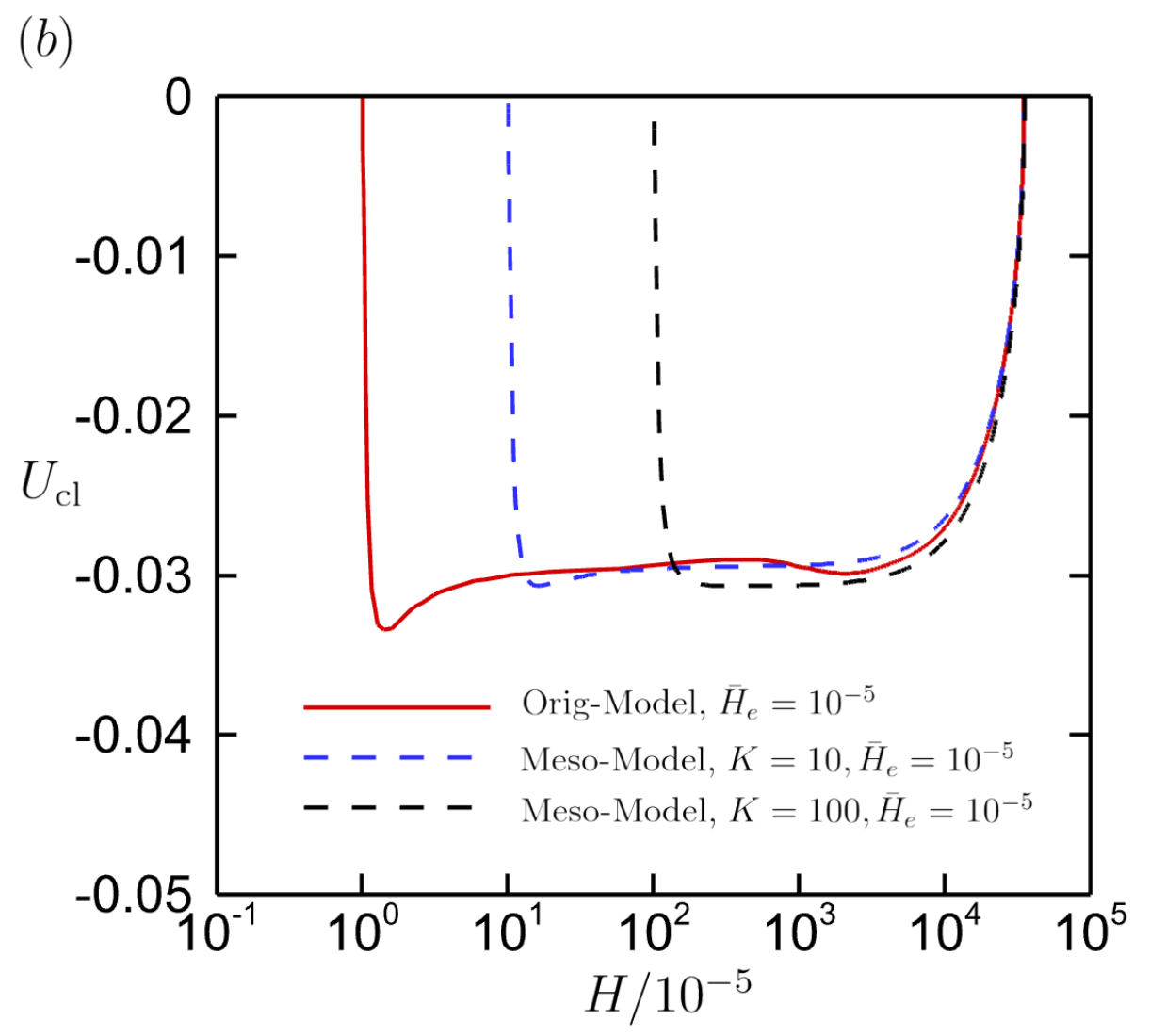}
  \caption{
  Profiles near the  contact line for axisymmetric drop retraction at a velocity of $-0.03$: (\textit{a}) distribution of the third power of the interface slope plotted against the distance to the  receding contact line with a logarithmic $X$-axis, where the legend is identical to that in Fig.~\ref{Fig7}; (\textit{b}) extended contact line velocity $U_{cl}$ as a function of the drop thickness $H \times 10^5$ (only the Orig-Model with $\bar H_e = 10^{-5}$ and the Meso-Model with $K = 10$ and $100$ are shown).
}
  \label{Fig9}
\end{figure}

Fig.~\ref{Fig7}(a) shows the temporal evolution of the macroscopic drop radius (the receding contact line position $X_{rec}$), and Fig.~\ref{Fig7}(b) depicts the drop interface profiles near the receding contact line at $T=1.0$. The distribution of the cubed interface slope $(H')^3$ is compared in Fig.~\ref{Fig9}(a) among the Meso-Models, the Orig-Models, and the Cox-Voinov theory at a uniform receding contact line velocity of $-0.03$. Furthermore, Fig.~\ref{Fig9}(b) illustrates the spatial variation of the extended contact line velocity $U_{cl}$, where a well-defined intermediate plateau is visible where $U_{cl}$ maintains a nearly constant value of approximately $-0.03$.

Similar to the spreading process, the macroscopic behaviors predicted by the Meso-Models agree remarkably well with the Orig-Models, as the Meso-Models faithfully reproduce the localized interfacial features within the intermediate region. Furthermore, the persistent constant plateau of $U_{cl}$ implies the invariance of the correction factor $\Theta$ in this region, also confirming the validity of the proposed Meso-Model under a receding contact line.

\subsection{Drop sliding on a plate}
\label{sec:slide}

The second example is the sliding of a drop on a plate driven by gravity, in which both advancing and receding contact lines simultaneously exist. Both 2D and 3D cases are considered here. This problem has been studied in experiments \cite{podgorski2001,peters2009} and numerical simulations \cite{benilov2015,solomenko2017,zhao2022}, where the plate can be inclined at some angle. Here, the plate is set to be vertical for simplicity. Consequently, gravity is incorporated by defining the potential as $\Phi = -Bo X$, where the Bond number $Bo = \rho g R^2 / (\sigma \bar \theta_e)$ represents the relative importance of gravitational effects over capillary forces.

In the simulations, the drop is set to slide from rest. The initial drop profile is prescribed as a hydrostatic cap without gravity, i.e.,
\begin{subequations}
\label{app:initslide}
\begin{align}
&H(X, 0) = \max \left( \frac{1-X^2}{2}, H_e \right), \quad \text{for the 2D case},\\
&H(X, 0) = \max \left( \frac{1-X^2-Y^2}{2}, H_e \right), \quad \text{for the 3D case}.
\end{align}
\end{subequations}

\subsubsection{2D drop sliding}
\label{sec:slide_2d}

We first consider the 2D drop sliding case. Under gravity, the drop begins to slide down and eventually reaches a constant steady-state velocity $U_s$. Following the same parameter configurations as in Section \ref{sec:asym}, five sets of simulations are performed using both the Orig-Models and the Meso-Models for comparison.

\begin{figure}
  \centering
  \includegraphics[scale=0.4]{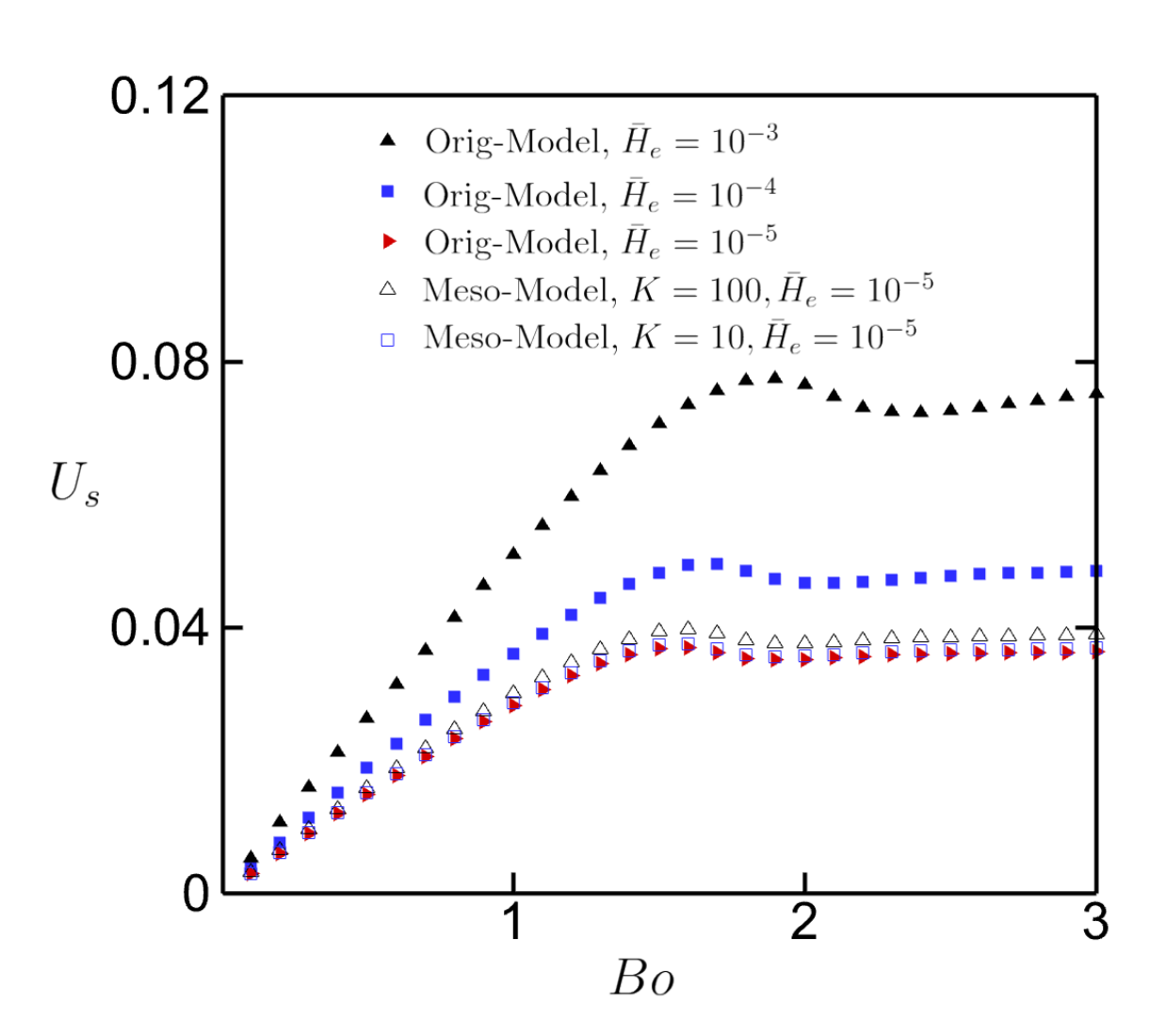}
  \caption{
2D drop sliding: steady-state sliding velocity $U_s$ plotted against the Bond number $Bo$.
}
  \label{Fig10}
\end{figure}

\begin{figure}
  \centering
  \includegraphics[scale=0.4]{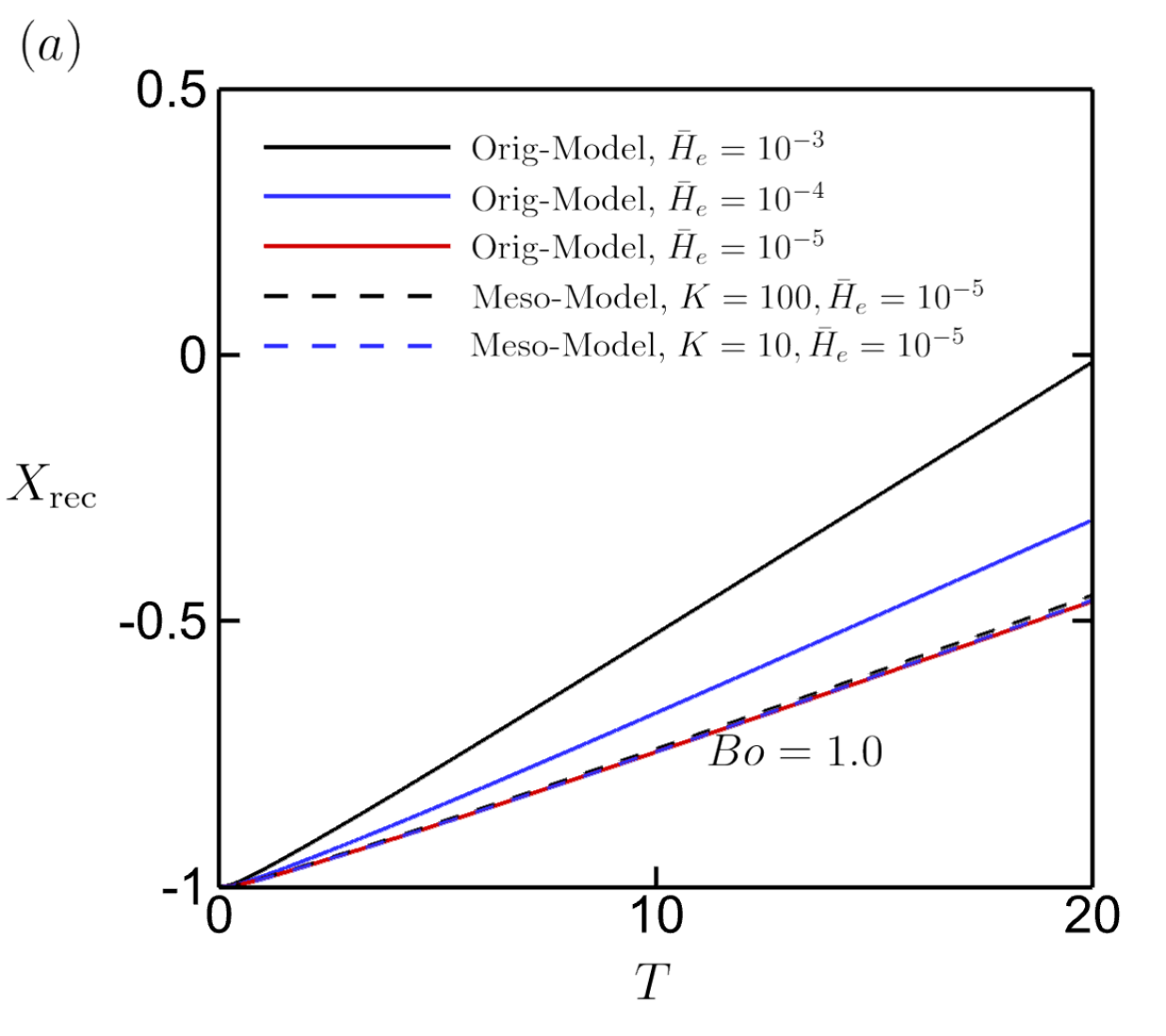}\includegraphics[scale=0.4]{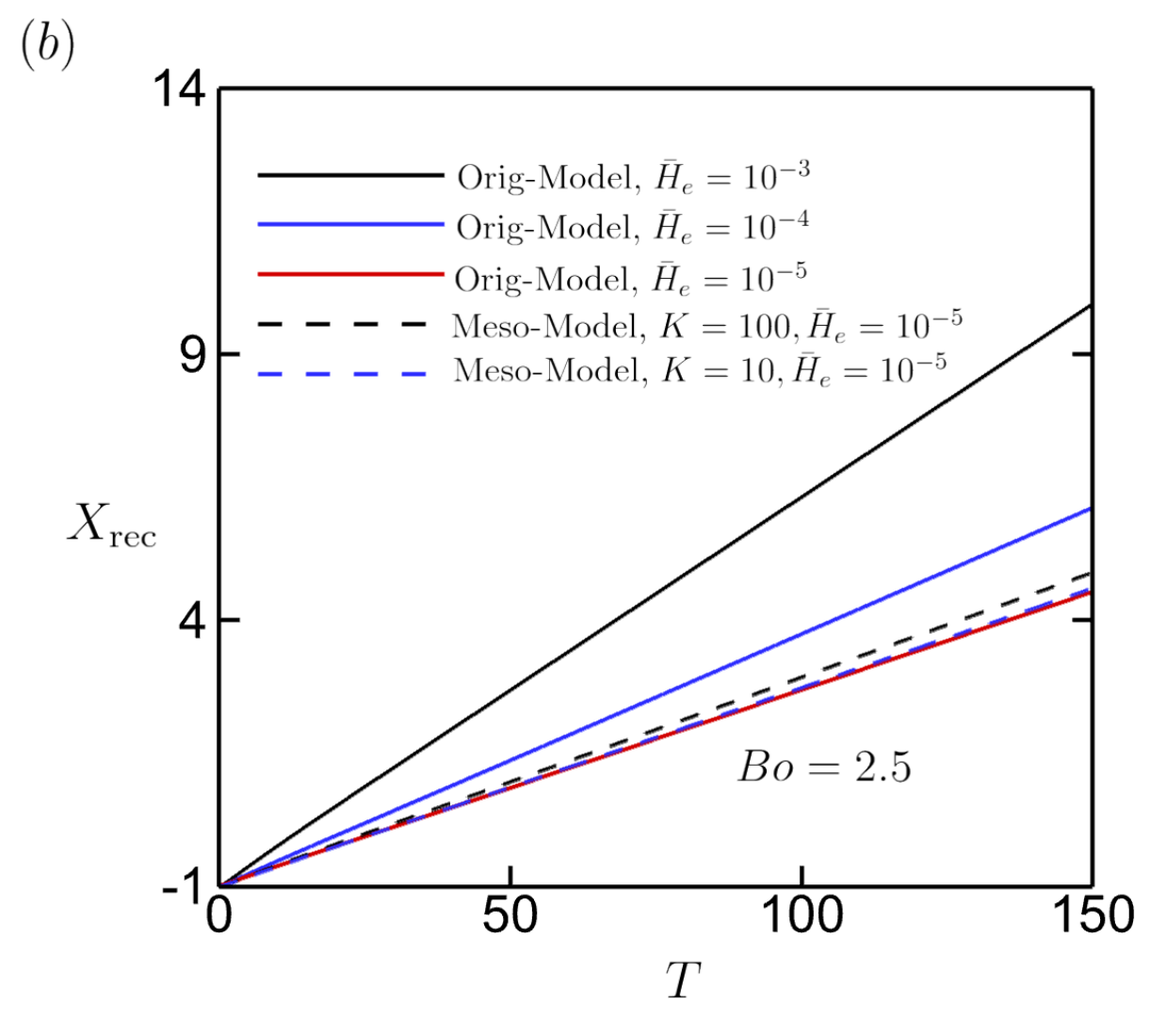}
  \caption{
 Temporal evolution of the  receding contact line position $X_{rec}$ for 2D drop sliding: (\textit{a}) $Bo = 1.0$ and (\textit{b}) $Bo = 2.5$.
}
  \label{Fig11}
\end{figure}

\begin{figure}
  \centering
  \includegraphics[scale=0.4]{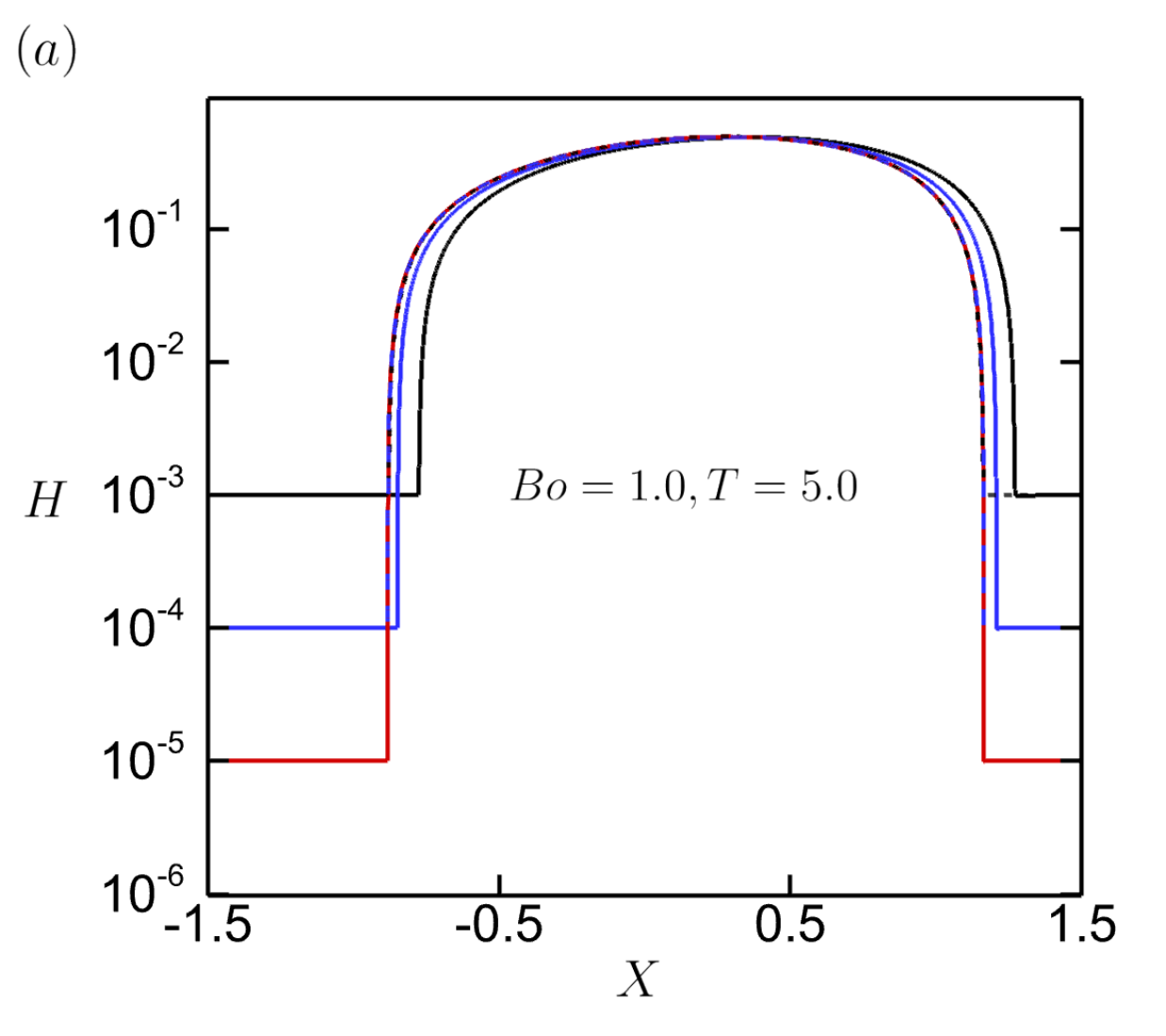}\includegraphics[scale=0.4]{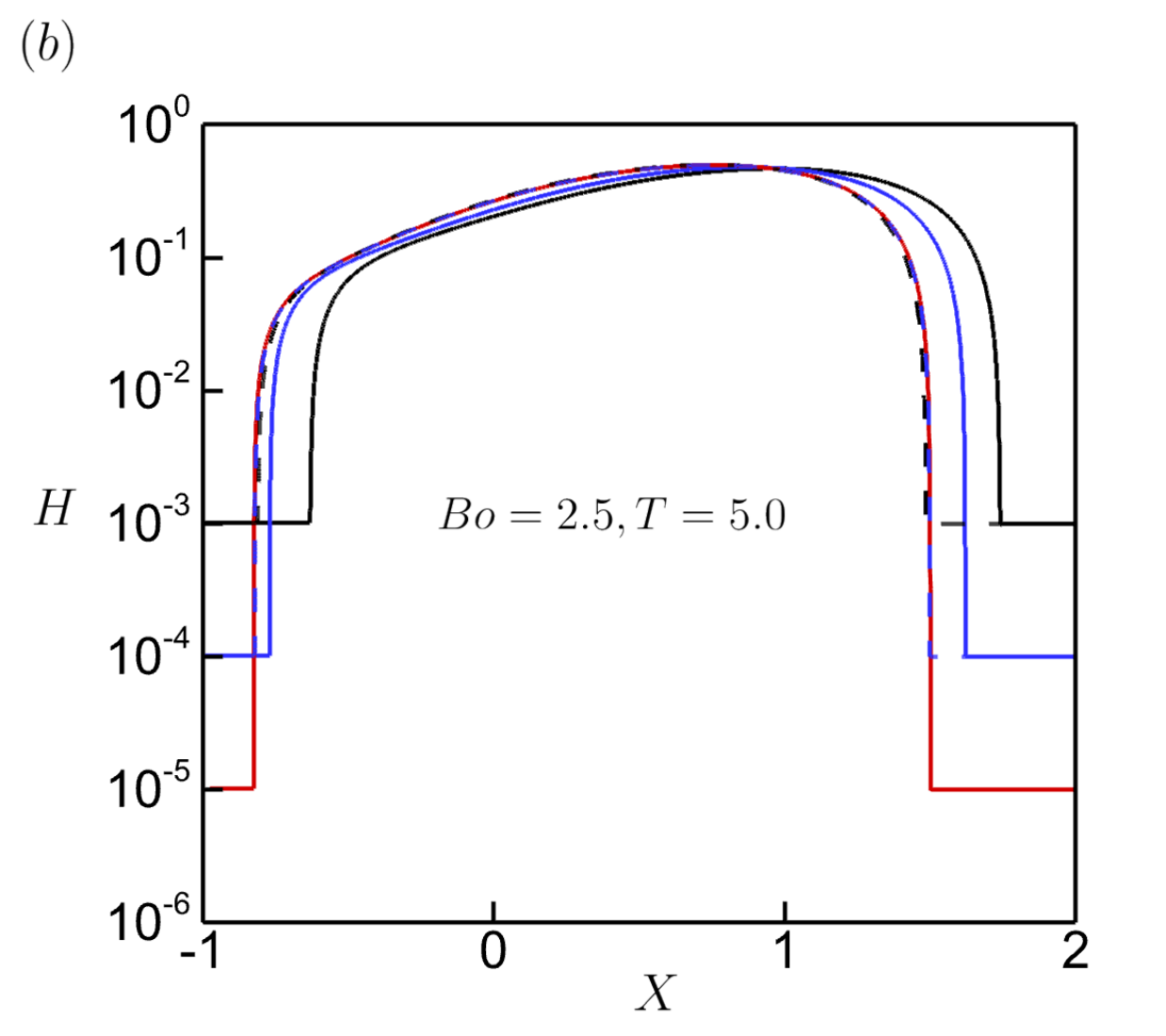}
  \caption{
  Drop interface profiles at $T = 5.0$ plotted with a logarithmic $H$-axis for 2D drop sliding: (\textit{a}) $Bo = 1.0$ and (\textit{b}) $Bo = 2.5$. The legend is identical to that in Fig.~\ref{Fig11}.
}
  \label{Fig12}
\end{figure}

Fig.~\ref{Fig10} shows the steady-state sliding velocity $U_s$ as a function of $Bo$ for $0 \le Bo \le 3.0$. For the Orig-Models, a thick precursor film ($\bar H_e = 10^{-3}$) severely overestimates the steady-state sliding velocity. As $\bar H_e$ decreases to $10^{-5}$, $U_s$ reduces significantly and converges toward the physical limit. For two typical Bond numbers, $Bo = 1.0$ and $Bo = 2.5$, Fig.~\ref{Fig11} presents the temporal evolution of the  receding contact line position $X_{rec}$, and Fig.~\ref{Fig12} presents the drop interface profiles with a logarithmic $H$-axis at $T = 5.0$. Due to the axisymmetric action of gravity, an elongated tail has gradually formed near the receding contact line in Fig.~\ref{Fig12}. The Meso-Models successfully captures the interface behavior of the Orig-Models within the intermediate region despite using a thicker precursor film.

Figs.~\ref{Fig10}-\ref{Fig12} indicate that both the steady-state and unsteady macroscopic behaviors predicted by the Meso-Models agree well with those from the corresponding Orig-Models. However, as $Bo$ increases, the deviation among the Orig-Model and corresponding Meso-Models gradually expands. This discrepancy is attributed to the use of the approximation \eqref{approx}, whose accuracy diminishes as the contact line velocity increases.

\subsubsection{3D drop sliding}

For the 3D sliding case, due to symmetry, only half of the drop domain is simulated to reduce the computational cost. Nevertheless, due to the relatively large computational cost, the Orig-Models with $\bar H_e = 10^{-5}$ or $\bar H_e = 10^{-4}$ are not simulated here. Instead, three Meso-Models with the parameter configurations as $K=100$, $K=200$ and $K=300$, are carried out to approximate the Orig-Model with $\bar H_e = 10^{-5}$ and to check if their results are consistent. Additionally, the unphysical Orig-Model with $\bar H_e = 10^{-3}$ is also simulated for comparison.

As reported by Ref. \cite{podgorski2001}, a critical Bond number exists beyond which a wetting transition occurs. Before this transition, a sharp corner typically emerges at the rear of the drop. Furthermore, at high Bond numbers, the sliding drop undergo a pinch-off process during its evolution. Consequently, we select $Bo = 2.0$ (before the wetting transition) and $Bo = 4.0$ (where both transition and pinch-off occur) as two representative cases to test our models.

\begin{figure}
  \centering
  \includegraphics[scale=0.4]{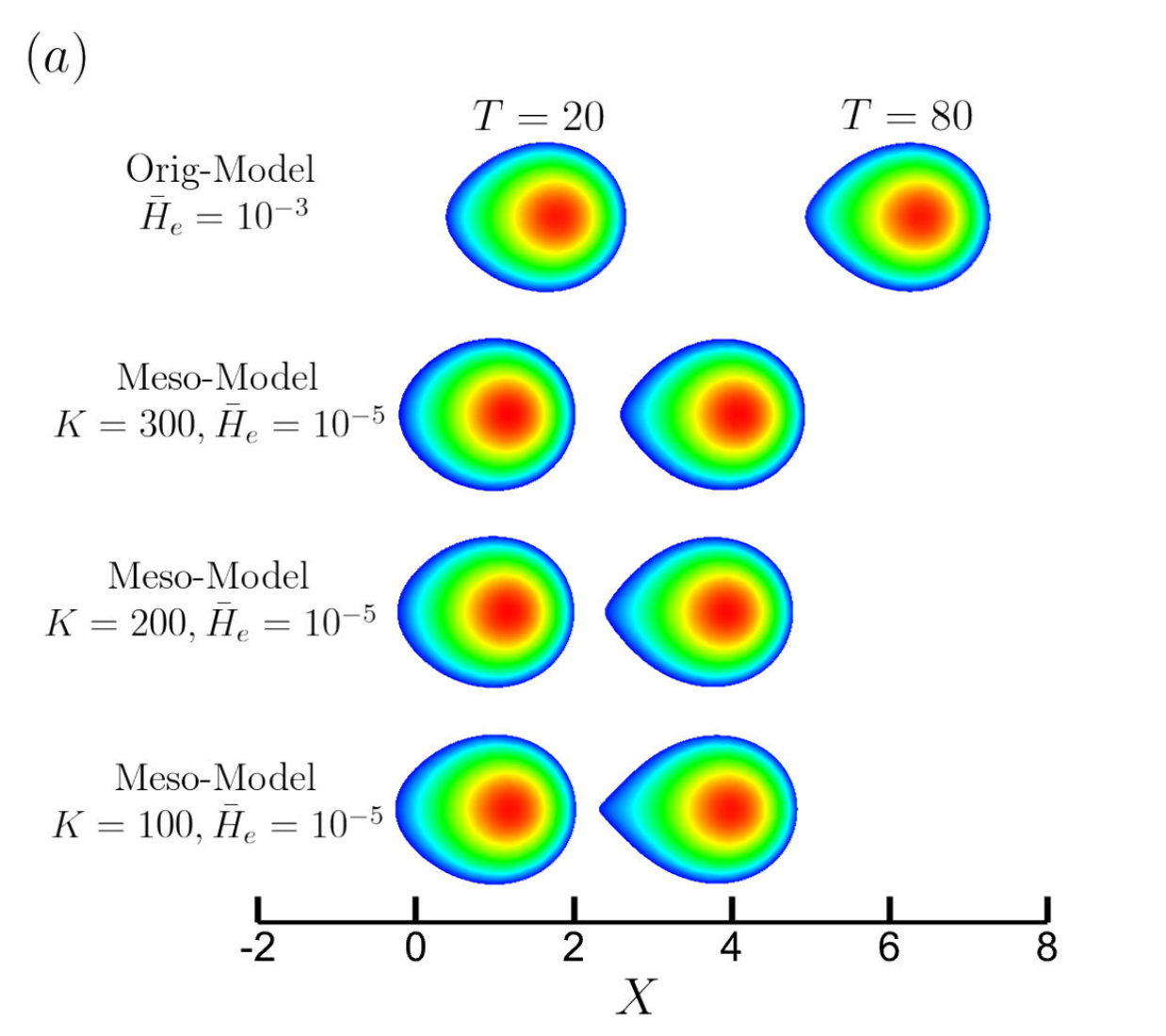}\includegraphics[scale=0.4]{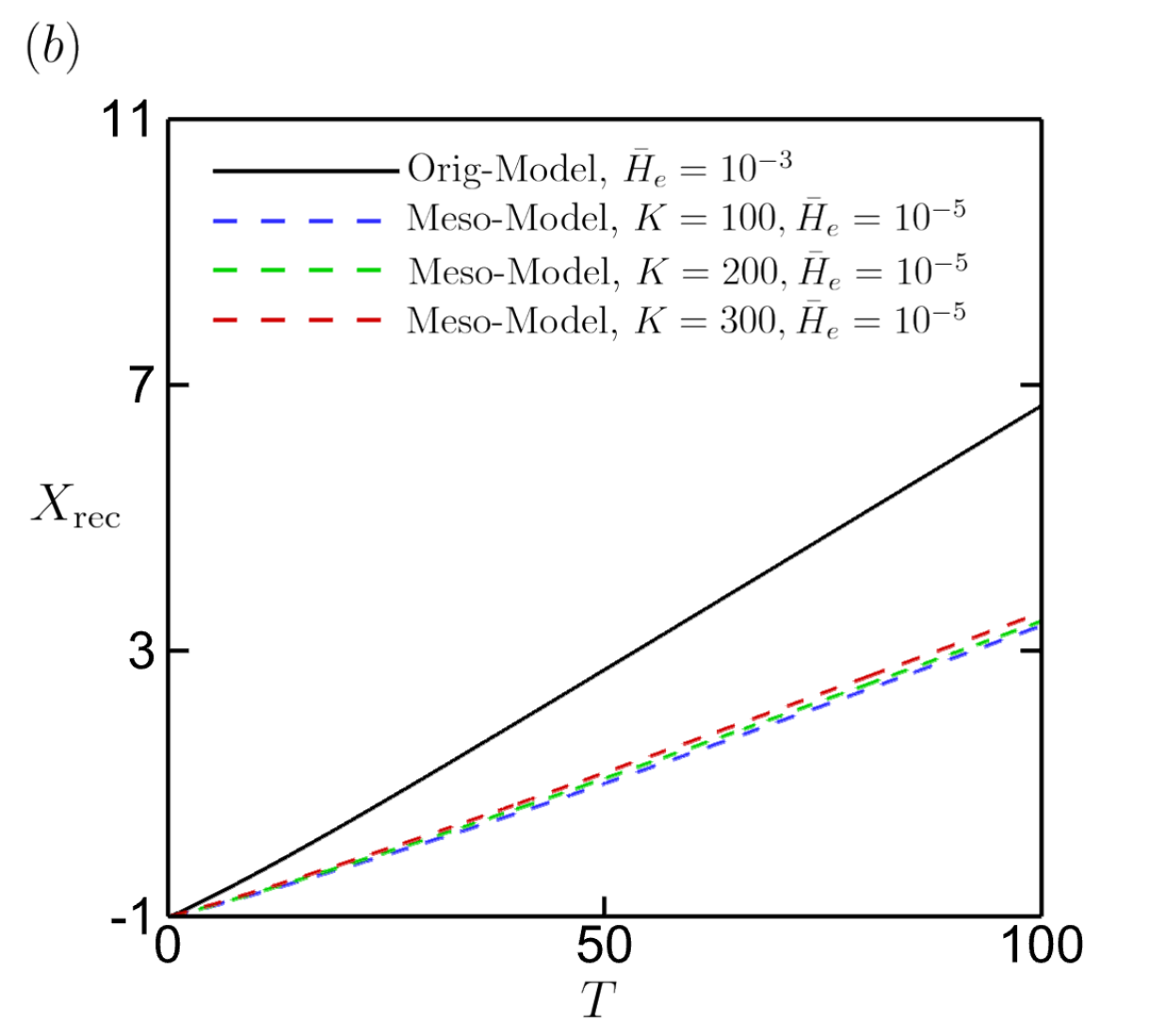}
\caption{
Comparison between the Orig-Model and the Meso-Models (with different scaling ratios $K$ all targeting $\bar H_e = 10^{-5}$) for 3D drop sliding at $Bo = 2.0$: (\textit{a}) drop contours (with the precursor film truncated) at $T = 20$ and $T = 80$; (\textit{b}) temporal evolution of the  receding contact line position $X_{rec}$.
}
  \label{Fig13}
\end{figure}

\begin{figure}
  \centering
  \includegraphics[scale=0.3,trim=0cm 0cm 1cm 0cm,clip]{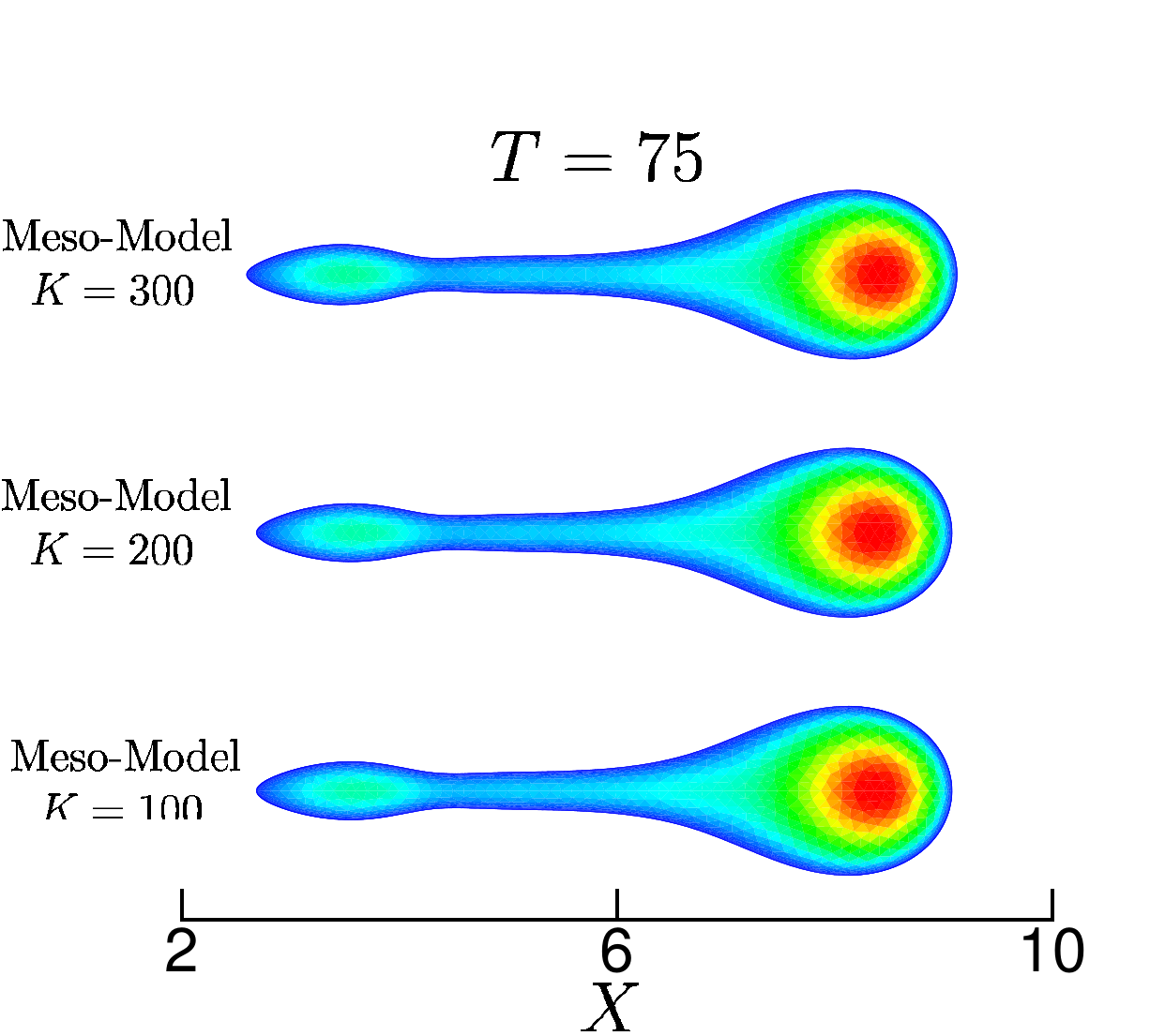}\includegraphics[scale=0.3,trim=2cm 0cm 1.5cm 0cm,clip]{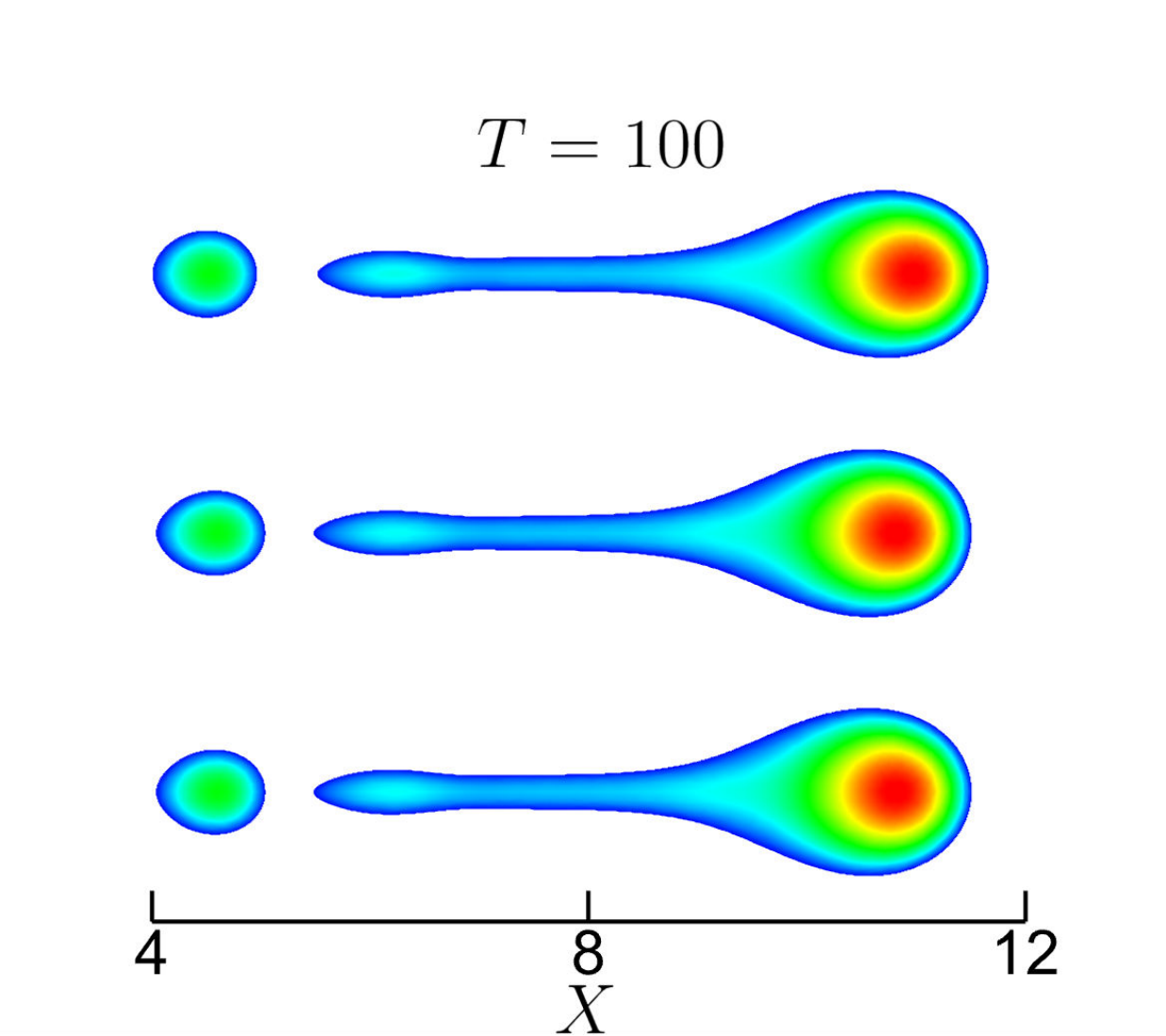}\includegraphics[scale=0.3,trim=2cm 0cm 1.5cm 0cm,clip]{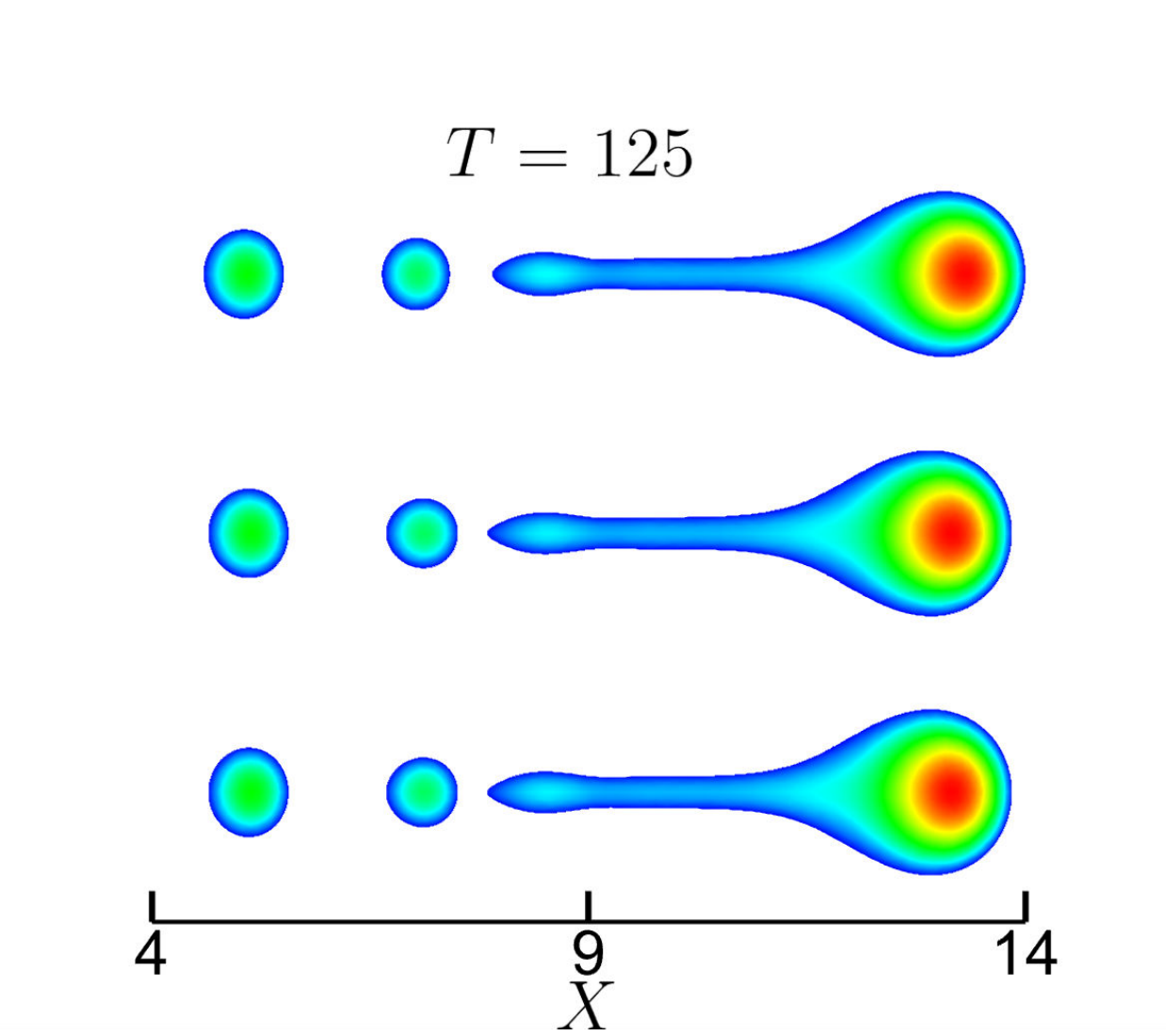}
  \caption{
  Comparison of drop contours (with the precursor film truncated) for 3D drop sliding obtained from the Meso-Models with different scaling ratios $K$ (all targeting $\bar H_e = 10^{-5}$): (\textit{a}) $T = 75$; (\textit{b}) $T = 100$; (\textit{c}) $T = 125$.
  }
  \label{Fig14}
\end{figure}

For the case of $Bo=2.0$, comparisons among the Orig-Model and the Meso-Models are presented in Fig.~\ref{Fig13}. Specifically, Fig.~\ref{Fig13}(a) displays the drop contours at $T = 20$ and $T = 80$, while Fig.~\ref{Fig13}(b) shows the temporal evolution of the  receding contact line. As shown in Fig.~\ref{Fig13}(a), while the unphysical Orig-Model with $\bar H_e = 10^{-3}$ severely deviates from the others, the temporal evolution results from the three Meso-Models show good consistency. In Fig.~\ref{Fig13}(a), the sliding drops have all reached a constant steady-state velocity at $T = 80$. At this stage, the drop tail predicted by the Meso-Model with $K=100$ is slightly sharper than those from the other two Meso-Model configurations. Ref. \cite{solomenko2017} indicated that such a corner profile can be captured numerically only when a physically small slip length is adopted. Furthermore, Refs. \cite{eggers2005a,sibley2015} demonstrated that slip models and precursor film models are macroscopically equivalent. Therefore, it can be inferred that the sharp corner profile can be resolved only if a physically small precursor film thickness is utilized, which is successfully achieved here by the present Meso-Models. Although the Meso-Model with $K=100$ is considered to be more accurate, the discrepancy produced by the other two Meso-Models remains within an acceptable range.

For the case of $Bo=4.0$, temporal comparisons among the three Meso-Models are presented in Fig.~\ref{Fig14}. At $T = 75$, the drop tail elongates and forms a sharp corner. As time progresses to $T = 100$, a pinch-off event occurs at the rear, shedding a satellite drop. At $T = 125$, the drop neck further breaks up, leaving behind two satellite drops. The three Meso-Models show good consistency, validating the capability of the Meso-Model in handling complex topological changes such as drop pinch-off.

\subsection{Drop coalescence}
\label{sec:coalescence}

The third example is the coalescence of two drops resting on a flat substrate without any external forces ($\Phi=0$), which has been studied both experimentally \cite{hernandez2012, eddi2013} and numerically \cite{eggers1999,eggers2025}. The initial film thickness profile $H(\mathbf{X},0)$ is defined as
\begin{equation}
H(\mathbf{X},0) = \max \left( \frac{R_0^2 - [ X^2 + (Y+0.95)^2 ]}{2R_0^4}, \frac{R_0^2 - [X^2 + (Y-0.95)^2 ]}{2R_0^4}, H_e \right),
\end{equation}
where the initial drop radius is set to $R_0 = 0.9<1$. Driven by capillarity, the two drops will spontaneously spread and subsequently undergo coalescence.

\begin{figure}
  \centering
  \includegraphics[scale=0.6,trim=0cm 6cm 0cm 6cm,clip]{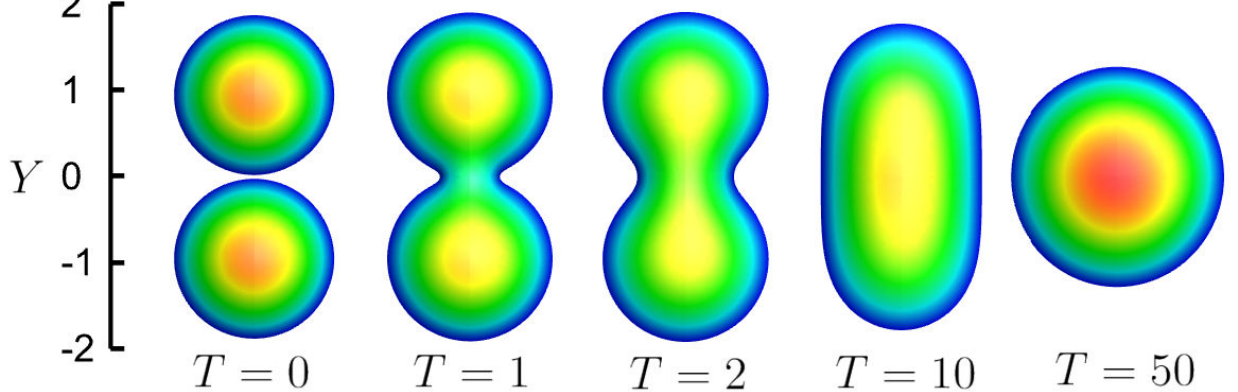}
  \caption{
  Drop contours (with the precursor film truncated) for drop coalescence obtained from the Meso-Model with $H_e = 3\times10^{-3},~K=300$ at $t = 0$, $1$, $2$, $10$ and $50$.
  }
  \label{Fig15}
\end{figure}

We conducted numerical simulations for target thin precursor film thickness $\bar H_e = 10^{-5}$ using three different Meso-Models with scaling ratios $K = 100, 200$ and $300$. The results from these three configurations are mutually consistent and show good convergence; therefore, for clarity, only the temporal evolution obtained with $K = 300$ (where $H_e = 3 \times 10^{-3}$) are presented in Fig.~\ref{Fig15}. At the beginning ($T = 0.0$), two identical, separate drops are initially deposited in close proximity on the substrate. As time progresses, the two drops make contact with each other, and their contact lines start to merge. Eventually, at $T = 50.0$, the fluid merges into a single larger circular drop. This dynamic process demonstrates the capability of the proposed Meso-Model in handling complex topological changes during drop coalescence.

\subsection{Breakup of liquid ridge}
\label{sec:breakup}

The fourth example considers the breakup of a liquid ridge on a flat substrate without any external forces ($\Phi=0$). Previous studies indicate that stationary ridges are susceptible to the Rayleigh-Plateau instability \cite{king2006, diez2009}. This problem has been investigated both experimentally and numerically, showing that a liquid ridge will spontaneously break up and shed smaller satellite droplets \cite{peschka2019, peschka2022}.

\begin{figure}
  \centering
  \includegraphics[scale=0.6,trim=0cm 0cm 0cm 0cm,clip]{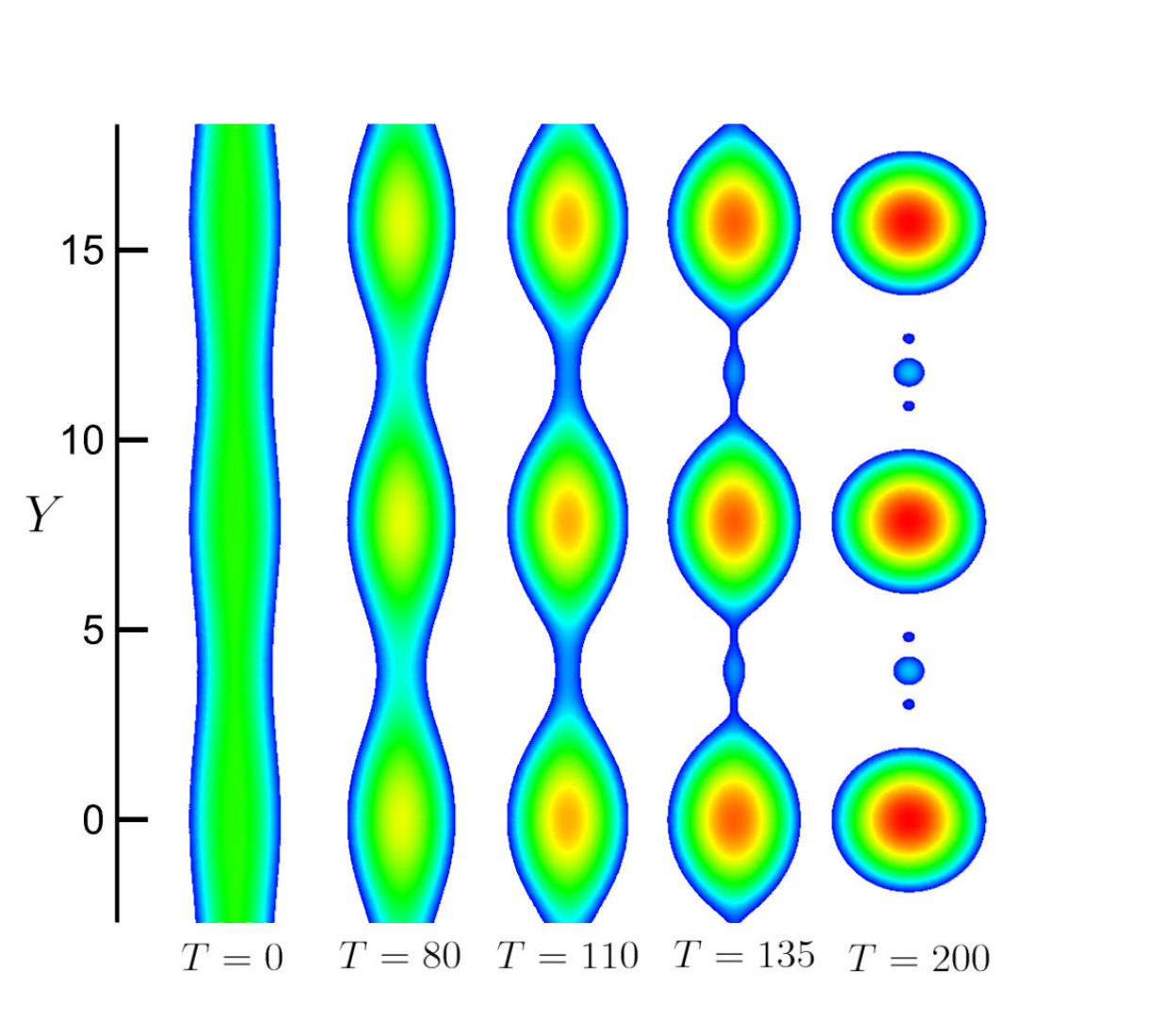}
  \caption{
  Evolution of morphology (with the precursor film truncated) during the breakup of a liquid ridge, obtained from the Meso-Model with $H_e = 3\times10^{-3}$ and $K=300$ at $T = 0$, $80$, $110$, $135$, and $200$.
  }
  \label{Fig16}
\end{figure}

The initial film thickness profile $H(\mathbf{X},0)$ is constructed by adding a perturbation along the $Y$-direction to a liquid ridge:
\begin{equation}
H(\mathbf{X},0) = \max \left( \frac{1}{2} - \frac{[X - 0.1 \cos(\lambda Y)]^2}{2}, H_e \right),
\end{equation}
where the perturbation wavenumber is set to $\lambda = 0.8$. The computational domain is restricted to $[0, 2.5] \times [0, 2\pi/\lambda]$ by applying Neumann boundary conditions on all boundaries. Due to this setup, the full numerical results can be obtained by mirroring the solution across the $XZ$- and $YZ$-planes.

We conducted numerical simulations using three different Meso-Models with scaling ratios $K = 100, 200$, and $300$. The results from these three configurations are mutually consistent and show good convergence; therefore, for clarity, only the results obtained with $K = 300$ (where $H_e = 3 \times 10^{-3}$) are presented in Fig.~\ref{Fig16}. Driven by the Rayleigh-Plateau instability, the liquid ridge undergoes significant spanwise dynamic deformation, where fluid gradually drains from the thinner neck regions into the bulging regions, as shown at $T = 80$ and $T = 110$. By $T = 135$, the neck region pinches off, leading to the breakup of the liquid ridge. During this pinch-off process, a series of satellite droplets are shed, forming a distinct droplet cascade structure that is clearly visible at $T = 200$. The formation of satellite droplets and the cascaded morphology show good agreement with both the numerical simulations and experimental observations reported in \cite{peschka2019}.

\begin{table}[width=.9\linewidth,cols=4,pos=h]
\caption{Comparison of computational costs (in core-hours) among different configurations of Meso-Models or Orig-Models for various examples. The recorded core-hours are approximate rounded values.}\label{tbl1}
\begin{tabular*}{\tblwidth}{@{} CCCC@{} }
\toprule
2D problems & $\bar H_e=1\times 10^{-5}$ & $H_e=1\times 10^{-4},~K=10$ & $H_e=1\times 10^{-3},~K=100$ \\
\midrule
Axisymmetric drop spreading & 1000 & 120 & 5 \\
Axisymmetric drop retraction & 900 & 100 & 4 \\
2D drop sliding ($Bo=4$) & 1500 & 180 & 10 \\
\midrule
3D problems & $H_e=1\times 10^{-3},~K=100$ &$H_e=2\times 10^{-3},~K=200$ & $H_e=3\times 10^{-3},~K=300$\\
\midrule
3D drop sliding ($Bo=4$) & 300 & 60 & 40 \\
Drop coalescence & 40 & 12 & 8 \\
Breakup of liquid ridge & 80 & 20 & 14 \\
\bottomrule
\end{tabular*}
\end{table}

For the six numerical examples across different model configurations, Table~\ref{tbl1} summarizes the computational costs (measured in core-hours), and quantitatively demonstrate the remarkable enhancement in computational efficiency achieved by the proposed Meso-Models compared to the physically realistic thin precursor film model ($\bar H_e = 10^{-5}$). For the first three 2D examples, transitioning from the Orig-Model to the Meso-Model with $K = 100$ leads to a substantial reduction in computational cost, achieving an approximate 200-fold acceleration. For the 3D problems presented in the lower half of the table, a vertical comparison among the Meso-Models reveals that increasing the scaling ratio $K$ from 100 to 300 further slashes the computational overhead by approximately $70\%$. As verified in the above subsections, this acceleration is accomplished without sacrificing the accuracy of the macroscopic flow dynamics. Therefore, to maximize computational efficiency while maintaining high accuracy, the Meso-Model with $K = 300$ is highly recommended for practical simulations of dynamic wetting processes.

\section{Conclusions}
\label{sec:conc}

In this work, a finite element method has been presented within the framework of lubrication theory and the precursor film model to simulate dynamic wetting processes. By modifying the disjoining pressure, the proposed mesoscopic precursor film model enables a relatively thick precursor film to accurately reproduce the intermediate-region interfacial behavior of a physically thin precursor film, thereby keeping the macroscopic flow dynamics essentially unchanged. Through a series of 2D and 3D simulations, the accuracy and capability of the proposed model are validated by examples including the spreading, retraction, sliding, and coalescence of the drops, as well as the breakup of liquid ridges. The numerical results show good agreement with available exact solutions, asymptotic theories, and experimental observations. The adoption of a thicker precursor film significantly reduces the computational cost. To maximize computational efficiency while maintaining high accuracy, the Meso-Model
with $K = 300$ is highly recommended for practical simulations of dynamic wetting processes. The present method efficiently handles complex topological changes in wetting and dewetting problems, which the slip models fail to capture.

\section*{Declarations of Interests}
The authors report no conflict of interest.

\section*{Funding}
This work was supported by the NSFC (grant nos 12241204, 12325208 and 12388101) and China Postdoctoral Science Foundation (grant nos GZC20261072 and 2025M781801).

\appendix

\section{Active region of the  disjoining pressure modification factor $\Theta$}
\label{appendix1}

An axisymmetric drop spreading case is selected as the benchmark (see Appendix B for formulation details). By comparing the numerical results under different configurations of disjoining pressure modifications $\Theta$, its precise active region can be identified.

\begin{figure}
  \centering
  \includegraphics[scale=0.4]{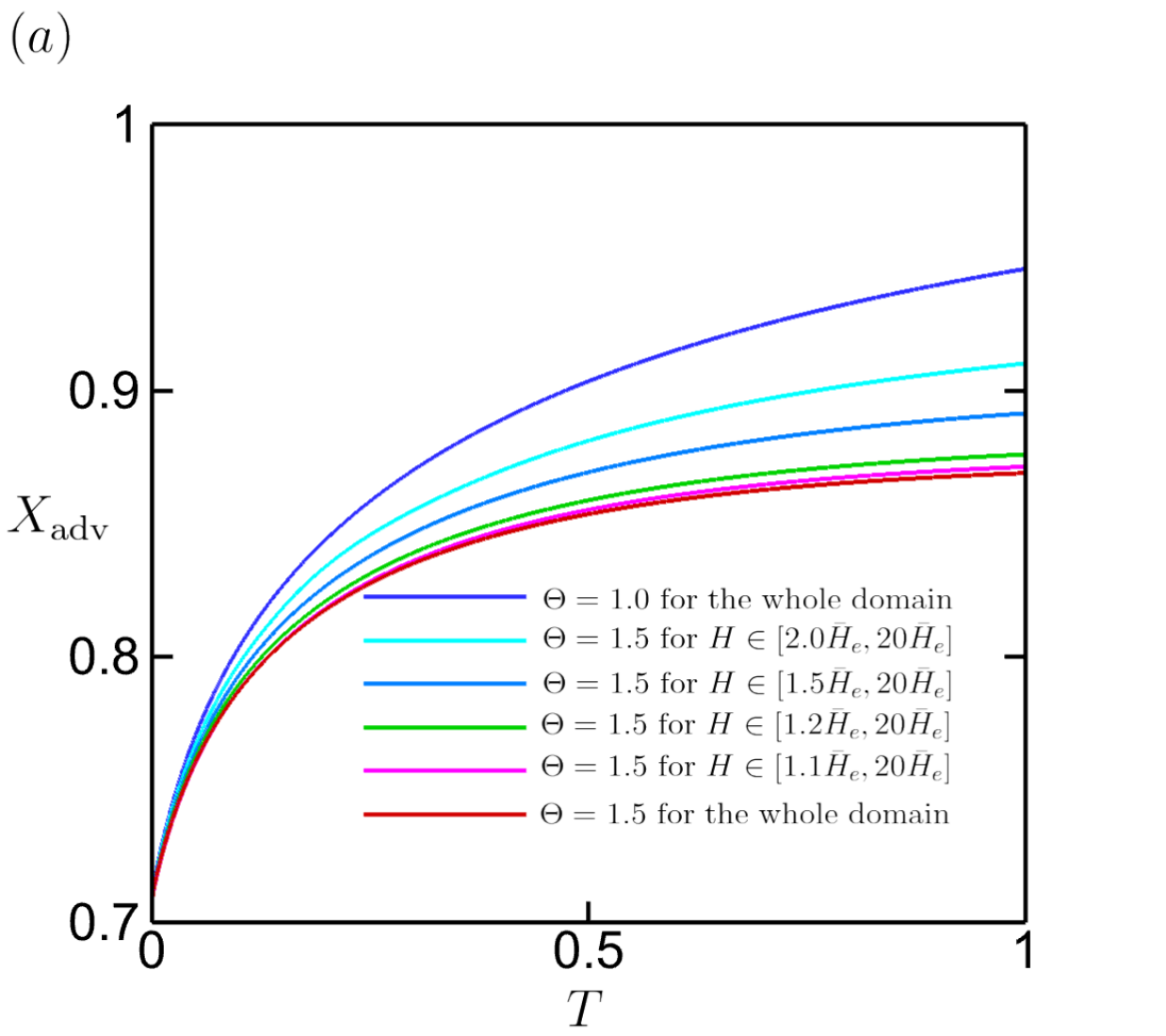}\includegraphics[scale=0.4,trim=0.1cm 0cm 0cm 0.1cm,clip]{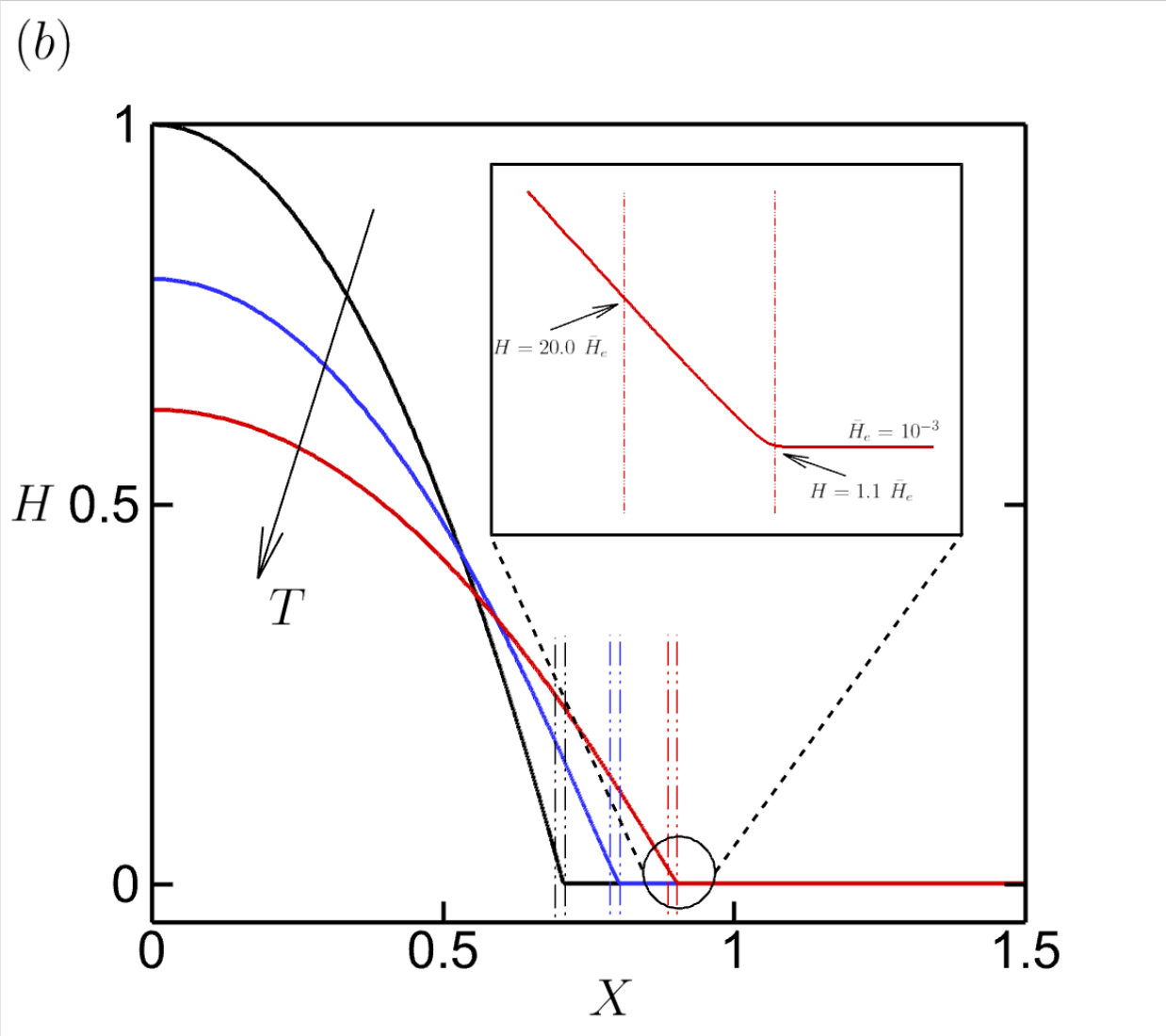}
  \caption{(\textit{a}) Temporal evolution of the dimensionless drop radius (advancing contact line position $X_{adv}$) during axisymmetric spreading under various active regions of $\Theta$.
  (\textit{b}) Dimensionless drop profiles $H(X)$ at different times, where the inset illustrates the spatial window of the localized disjoining pressure correction restricted within the thickness interval $H \in [1.0, 20.0] \bar H_e$.
}
  \label{Fig2}
\end{figure}

The initial drop radius is set to $R_0 = \sqrt{2}/2$, the precursor film thickness to $\bar H_e = 10^{-3}$, and the temporal step size to a constant value of $\Delta T = 10^{-4}$. Various configurations of $\Theta$ are prescribed to investigate their influence on the macroscopic drop evolution. Figure \ref{Fig2}(a) illustrates the temporal evolution of the drop radius (i.e., the  advancing contact line position). Here, $\Theta = 1$ corresponds to the baseline model with the original disjoining pressure, whereas $\Theta = 1.5$ represents the modified model that effectively mimics a thinner precursor film, thereby slowing down the drop spreading. By varying the active window of $\Theta = 1.5$, we find that the disjoining pressure correction plays a significant role only within a narrow, localized zone near the  contact line, specifically for $H \in [1.0, 20] \bar H_e$, as shown in Fig. \ref{Fig2}(b). Beyond this threshold, modifying the disjoining pressure exerts a negligible impact on the macroscopic flow dynamics. Moreover, the drop dynamics exhibit a high sensitivity to the lower bound of this localized region; as shown in Fig. \ref{Fig2}(a), even a minor shift in the lower bound from $1.1 \bar H_e$ to $1.5\bar H_e$ leads to a noticeable deviation.

In summary, the application of the proposed mesoscopic precursor film model requires a constant $\Theta$ within the localized zone $H \in [1.0, 20] \bar H_e$ near the contact line. Furthermore, any upward shift or numerical discontinuity at the lower bound of this interval should be avoided.

\section{Formulation of axisymmetric drop spreading/retracting}
\label{appendix2}

We consider the axisymmetric spreading and retraction of a drop on a flat substrate as the benchmark problem. Gravity is neglected for simplicity, giving $\Phi = 0$.

The equilibrium drop radius is set to 1 for convenience, corresponding to the choice of the characteristic lateral length scale:
\begin{equation}
R=\left(\frac{4V\bar\theta_e}{\pi}\right)^{1/3},
\end{equation}
where $V$ denotes the volume of the drop. Accordingly, the equilibrium drop bulk profile in the $X\text{--}Z$ plane is approximately given by $H=(1-X^2)/2$. The initial drop profile is prescribed as a circular paraboloid combined with a flat equilibrium precursor film:
\begin{equation}
H(X, 0) = \max \left( \frac{1}{2R_0^2} - \frac{X^2}{2R_0^4}, \, \bar H_e \right),
\label{app:init}
\end{equation}
where $R_0$ denotes the initial drop radius. The drop undergoes retraction when $R_0 > 1$, whereas it exhibits spreading when $R_0 < 1$.

The original, unregularized dimensionless axisymmetric lubrication equation is
\begin{equation}
\frac{\partial H}{\partial T}+\frac{1}{X}\frac{\partial }{\partial X}\left\{ XH^3\frac{\partial }{\partial X}\left[\frac{1}{X}\frac{\partial }{\partial X}\left(X\frac{\partial H}{\partial X}\right)-\frac{2 \bar H_e^2}{H^3}+\frac{2 \bar H_e^4}{H^5}\right] \right\}=0.
  \label{app:lub_axis}
\end{equation}
Due to axisymmetry, the computational domain is restricted to the interval $X \in [0, 1.5]$ in the first quadrant. Boundary conditions at the symmetry center ($X = 0$) and the far-field boundary ($X = 1.5$) are imposed as follows:
\begin{equation}
\frac{\partial H}{\partial X} = \frac{\partial}{\partial X} \left[ \frac{1}{X} \frac{\partial}{\partial X} \left( X \frac{\partial H}{\partial X} \right) \right] = 0.
\label{app:bc_center}
\end{equation}

Notably, due to the presence of a unidirectional contact line, the global vector field $\mathbf{n}$ within the Meso-Model numerical simulations simplifies to $\mathbf{n}=\mathbf{1}$.


\bibliographystyle{unsrt} 

\bibliography{cas-refs}

\end{document}